\documentclass{aastex61}

\usepackage{natbib}
\usepackage{longtable}
\received{}
\revised{}
\accepted{}
\submitjournal{ApJ}

\shorttitle{Damping mechanisms for transverse coronal waves}
\shortauthors{Montes-Sol\'{\i}s \& Arregui}

\begin{document}

\title{Comparison of damping mechanisms for transverse waves in solar coronal loops}

\correspondingauthor{Mar\'{\i}a Montes-Sol\'{\i}s}
\email{mmsolis@iac.es}

\author[0000-0003-3587-6443]{Mar\'{\i}a Montes-Sol\'{\i}s}
\affil{Instituto de Astrof\'{\i}sica de Canarias, E-38205 La Laguna, Tenerife, Spain}
\affil{Departamento de Astrof\'{\i}sica, Universidad de La Laguna, E-38206 La Laguna, Tenerife, Spain}
\author[0000-0002-7008-7661]{I\~nigo Arregui}
\affiliation{Instituto de Astrof\'{\i}sica de Canarias, E-38205 La Laguna, Tenerife, Spain}
\affiliation{Departamento de Astrof\'{\i}sica, Universidad de La Laguna, E-38206 La Laguna, Tenerife, Spain}




\begin{abstract}
We present a method to assess the plausibility of alternative mechanisms to explain the damping of magnetohydrodynamic (MHD) transverse waves in solar coronal loops. The considered mechanisms are resonant absorption of kink waves in the Alfv\'en continuum, phase-mixing of Alfv\'en waves, and wave leakage. Our methods make use of Bayesian inference and model comparison techniques. We first infer the values for the physical parameters that control the wave damping, under the assumption of a particular mechanism, for typically observed damping time-scales.  Then, the computation of marginal likelihoods and Bayes factors enable us to quantify the relative plausibility between the alternative mechanisms. We find that, in general, the evidence is not large enough to support a single particular damping mechanism as the most plausible one. Resonant absorption and wave leakage offer the most probable explanations in strong damping regimes, while phase mixing is the best candidate for weak/moderate damping. When applied to a selection of 89 observed transverse loop oscillations, with their corresponding measurements of damping times scales and taking into account data uncertainties, we find that only in a few cases positive evidence for a given damping mechanism is available.    

\end{abstract}




\section{INTRODUCTION} \label{sec:intro}
	 
	The damping of magnetohydrodynamic (MHD) waves in solar coronal structures is a commonly observed phenomenon and a source of information about their physical conditions, dynamics, and energetics. The study of the damping of transverse waves has attracted particular attention, since the first imaging observations of decaying transverse coronal loop oscillations by \cite{Aschwanden1999} and \cite{Nakariakov1999}. High resolution imaging and spectroscopic observations with ground- and space-based observatories such as Hinode, CoMP, SDO/AIA, STEREO, Hi-C, and IRIS have enabled us to measure transverse wave dynamics with increasing precision in almost all layers of the solar atmosphere \citep[see e.g.,][]{okamoto11,White2012,Verwichte2013,morton13a,thurgood14,morton15,okamoto15}. Recent observational analyses have permitted the creation of databases containing the oscillation properties for a large number events and with the inclusion of measurement errors \citep{Verwichte2013,Goddard2016}.  The increase in the number of measured events and their properties including their uncertainty has led to advances in statistical seismology of coronal loops. \cite{Verwichte2013} considered a statistical approach to obtain information on the loop cross-sectional structuring parameters by forward modelling of scaling laws between periods and damping times, showing that restrictions can be found to the loop's density contrast and inhomogeneity layer. Following Bayesian methods, a number of studies by \cite{Arregui2011,Arregui2013a,Arregui2013b,Asensioramos2013,Arregui2014,Arregui2015} have shown how information on the uncertainty of the inferred coronal loop parameters and the plausibility between alternative models can be assessed.
	
	Although damping is not an omnipresent phenomenon, see e.g., \cite{anfinogentov2013,anfinogentov2015} for examples of decay-less oscillations or \cite{wang12} for an example of growing oscillations, theoretical explanations for the physical origin of the damping of transverse waves are abundant. Because direct viscous and resistive difussion time scales are too long in uniform plasmas, mechanisms based on the cross-field or field-aligned inhomogeneity of the wave guides become relevant. In the context of coronal loop oscillations, the discussion rapidly focused on mechanisms such as resonant absorption \citep{Goossens2002,Ruderman2002,Goossens2006}, phase mixing of Alfv\'en waves \citep{Priest1983}, lateral wave leakage \citep{Spruit1982,Cally1986,Roberts2000,Cally2003} or foot-point leakage at the chromospheric density gradient \citep{Berghmans1995,Depontieu2001,Ofman2002a}. Methods to assess the plausibility between alternative damping mechanisms are discussed in \cite{Roberts2000, Ruderman2005,Nakariakov2005,Aschwanden2005}. 
	
	Rough estimates of damping time scales predicted by the mechanisms under consideration, for typical coronal loop physical properties, and their comparison to observed damping time scales point to resonant damping as the most plausible mechanism, with phase mixing and wave leakage producing damping time scales that seem to be too long in comparison to those observed. The drawback to this method is the difficulty in defining what a typical coronal loop is, since the physical parameters of observed loops cannot be directly measured. 
	
	The modeling and analysis of wave properties for equilibrium states that enable the simultaneous occurrence of more than one mechanism is another alternative. The studies by e.g., \cite{Terradas2006}, \cite{Rial2013} and \cite{Soler2009a} indicate that resonant absorption in the Alfv\'en continuum is a far more efficient mechanism than lateral wave leakage in curved loops or than damping in the slow continuum in prominence threads. Resonant damping is frequency selective,  with low-frequency waves being favored in front of high frequency waves \citep{Terradas2010}.  A comparison between the outward to inward power ratio measurements in CoMP observations of coronal waves by \cite{Tomczyk2007} and the theoretical modeling by \cite{Verth2010} gives additional support to the resonant damping model.
	
	\cite{Ofman2002} proposed a method based on the use of scaling laws between the observed periods and damping times. Their suggestion is based on the assumption that each damping mechanism is characterized by a particular power law, with a distinct power index between the damping time and the oscillation period. By fitting the observed time-scales to those predicted by each damping mechanism, one can compare the theoretically predicted and fitted power indexes to discriminate between damping mechanisms. The application of this method has given results that support both phase mixing \citep{Ofman2002} and resonant absorption \citep{White2012,Verwichte2013}. It was pointed out by \cite{Arregui2008} that the use of scaling laws to discriminate between damping mechanisms is questionable.  For example, the resonant damping model is able to produce data realizations for which different scaling laws with different indexes may be obtained. The major drawback to this method is the faultiness of the assignment of a single power index to a particular damping mechanism, even more so when the damping time depends on a number of loop parameters that might have an intrinsic variability and values that are highly uncertain.
	
	We adopt a different approach and consider Bayesian methods to compute the level of plausibility among alternative damping models, conditional on observed data and considering their uncertainty. Inference and model comparison methods are applied to three particular damping mechanisms: resonant damping in the Alfv\'en continuum; phase mixing of Alfv\'en waves; and wave leakage of the principal kink mode.  These damping models are selected so as to give continuity to the discussion initiated by \cite{Ofman2002}, but the methods here presented can be extended in the future to include additional damping models in a straightforward manner. The methods are first applied to hypothetical data and then to a sample of real observations from the databases compiled by \citet{Verwichte2013} and \citet{Goddard2016}.
	
	The layout of the paper is as follows. Section~\ref{sec:models} gives a description of the damping models being compared. In Section~\ref{sec:bayes}, we present our methodology, based on the use of Bayes' rule for parameter inference and model comparison. Our results are presented in Section~\ref{results}. We first apply Bayes' rule to the problem of inferring the relevant physical parameters, under the assumption that each of the considered damping models is true. Then, we assess the plausibility of each damping model conditional on both hypothetical and real observations of transverse loop oscillations. Our summary and conclusions are presented in Section~\ref{conclusions}.
		
	\section{DAMPING MODELS} \label{sec:models}
	In this work, we have studied the degree of plausibility of three damping mechanisms in explaining the observations of transverse loop oscillations. 
	The first considered damping model is resonant absorption in the Alfv\'en continuum. This mechanism consists of an energy transfer between the global kink mode of a magnetic flux tube to Alfv\'en oscillations at the boundary of the tube, due to the transverse variation of Alfv\'en velocity within a layer that separates the tube and the background corona. Resonant absorption has been studied extensively and in great detail and is known to be able to produce time and spatial damping scales comparable to those observed.
	
	Following \citet{Goossens2002} and \citet{Ruderman2002}, we focus on the fundamental kink mode of a cylindrical density tube of length L and radius R with a uniform magnetic field along the axis of the tube and a non-uniform variation of the cross-field density on a length-scale $l$. Under the thin tube ($L>>R$) and thin boundary ($l/R<<1$) assumptions the expression for the damping ratio is 
	\begin{equation}\label{eq1}
		\frac{\tau_d}{P}=\frac{2}{\pi}\frac{R}{l}\frac{\zeta+1}{\zeta-1},
	\end{equation}
	with $\tau_{d}$ the damping time, P the oscillation period, $l$ the thickness of the non-uniform layer, and $\zeta=\rho_{i}/\rho_{e}$ the density contrast between the internal ($\rho_{i}$) and external ($\rho_{e}$) densities. The factor $2/\pi$ is due to the assumption of a sinusoidal profile for the density at the non-uniform boundary layer.
		
	According to Equation~(\ref{eq1}), the damping ratio due to resonant absorption is a function of two parameters, $\zeta$ and $l/R$. As shown by \citet{Goossens2002}, considering a typical contrast of $\zeta=10$, the mechanism is able to explain observed damping time scales for values of $l/R$  in between 0.1 and 0.5. Considering values of $\zeta$ and $l/R$  within plausible ranges, $\zeta \in (1,10]$ and $l/R \in (0,2]$, the values for the damping ratio predicted by theory are in a wide range $\tau_{d}/P \thicksim (0.5 - 10^{4})$. The mechanism is therefore able to predict damping properties compatible with observations.

	The next considered mechanism is associated with the phase mixing of Alfv\'en waves, first discussed by \citet{Priest1983} in the context of coronal heating. The basic idea is that Alfv\'en waves propagating along the magnetic field in a medium with a transverse gradient of Alfv\'en velocity become rapidly out of phase. As time progresses, increasingly shorter spatial scales are created. The mechanism has been discussed as a possible reason to explain the observed damping in \citet{Roberts2000, Nakariakov2005} and \citet{Ofman2002}. In his analysis, \citet{Roberts2000} obtained an analytical expression for the damping ratio of the form 
	\begin{equation}\label{eq2}
		\frac{\tau_d}{P}=\left (\frac{3}{\pi^2 \nu} \right) ^{1/3} w^{2/3}P^{-1/3},
	\end{equation}
	where $\nu=4\times 10^{3}$ km$^2$ s$^{-1}$ is the coronal kinematic shear viscosity coefficient and $w$ the transverse inhomogeneity length scale. Note that in Equation (\ref{eq2}), the damping ratio is a function of the period. Considering a period in the range $P\in[150,1250]$s and $w \in [0.5,20]$Mm, damping ratios in the range $\tau_{d}/P \thicksim [0.3 - 6]$ are obtained, so this mechanism could also explain the observed damping time scales, although it has serious limitations of applicability to transverse loop oscillations. \cite{Ofman2002} considered phase mixing as a mechanism involving the friction between adjacent unresolved thin loop strands, a hypothesis that remains to be demonstrated. The mechanism is here included in the interest of continuity in the discussion by considering the same damping mechanisms discussed by \cite{Ofman2002}, but presenting an alternative method to the scaling law approach for their comparison.

	The third considered mechanism is wave leakage of the principal leaky mode, as discussed by \citet{Cally2003} and \citet{Terradas2007}. This theoretical solution consists of a radiating wave which oscillates with the kink mode frequency and looses part of its energy to the background corona. Following \citet{Cally2003}, we consider the analytic approximate solution  for the damping ratio in the thin tube limit and valid for $\rho_i>>\rho_e$ of the form 
	\begin{equation}\label{eq3}
		\frac{\tau_d}{P}=\frac{4}{\pi^4} \left (\frac{R}{L}\right)^{-2}.
	\end{equation}
	
	Considering values of R/L in the plausible range R/L$\in[10^{-4},0.3]$, theory predicts damping ratios in the range $\tau_{d}/P \thicksim(0.5 - 10^5)$.

	 For simplicity, strong assumptions are made in the adopted damping mechanisms and formulas. The resonant damping formula assumes the thin tube and thin boundary approximation and the adoption of a particular radial density profile. The influence of considering different density profiles in the inferences using resonant absorption was analyzed by \citet{Arregui2015}. This influence was found to be important in strong damping regimes. The phase-mixing damping formula is valid in the strongly developed regime and the value of kinematic viscosity is kept constant. For wave-leakage, the possible influence of the density contrast was ignored. These assumptions lead to the existence of hidden variables that may vary from loop to loop. All loops do not necessarily have the same radial density profile, temperature, viscosity and density contrast.
		
	Once the three damping mechanisms to be considered in this study are presented, it is worth clarifying that they are going to be compared on the basis of their ability to reproduce (in a statistical way) the observed periods and damping time scales, taking into account those observed time scales and their associated uncertainties.
	
	\section{BAYESIAN METHODOLOGY}\label{sec:bayes}
	
	In this paper, we adopt the methods of Bayesian analysis to perform parameter inference and model comparison. Bayesian reasoning goes back to the essay by \cite{bayes63} and was given its current mathematical basis by \cite{laplace74}. It offers a principled way to assess the plausibility of statements conditional on the available information. In recent years, it has become widely used to perform scientific inference in areas such as the physical sciences \citep{Vontoussaint11}, cosmology \citep{Trotta08} or exoplanet research \citep{Gregory05}.
	
	We have used Bayesian analysis tools to perform parameter inference, assuming a particular damping model is true, and to compare the relative plausibilities of the three damping models described above.
	
	\subsection{Parameter Inference}\label{inference}
	In the Bayesian framework the inference of a parameter set $\mbox{\boldmath$\theta$}$ that characterizes a model M conditional on observed data $d$ is performed by making use of the Bayes' theorem
	\begin{equation}\label{eq4}
		p(\mbox{\boldmath$\theta$}|M,d)=\frac{p(\mbox{\boldmath$\theta$}|M) p(d|M,\mbox{\boldmath$\theta$)}}{p(d|M)}.
	\end{equation}
	In this expression, $p(\mbox{\boldmath$\theta$}|M,d)$ is the posterior probability of the parameters conditional on the assumed model and the observed data; $p(\mbox{\boldmath$\theta$}|M)$ is their prior probability, before considering the data; and $p(d|M,\mbox{\boldmath$\theta$})$ is the likelihood function. The denominator in Equation (\ref{eq4}) is a normalization factor called marginal likelihood given the model M, integrated likelihood or model evidence. It represents the probability of the data given the model, and is computed by integrating the likelihood times the prior over the parameter space
	\begin{equation}\label{eq5}
		p(d|M)=\int_{\mbox{\boldmath$\theta$}}{p(\mbox{\boldmath$\theta$}|M)p(d|M,\mbox{\boldmath$\theta$})d\mbox{\boldmath$\theta$}}.
	\end{equation} 
	
	When interested in obtaining information on one particular parameter $\theta_{i}$ of model M, we compute its marginal posterior as
	\begin{equation}\label{eq6}
		p(\theta_i|M,d)=\int{p(\mbox{\boldmath$\theta$}|M,d)d\theta_1...d\theta_{i-1}d\theta_{i+1}...d\theta_n}, 
	\end{equation}   
	which is an integral of the full posterior over the rest of the n model parameters.
	
	\subsection{Model Comparison}
	The Bayesian framework also offers a straightforward method to compare the relative goodness of alternative models in explaining the observed data. Application of Bayes' theorem to N alternative models leads to
	
	\begin{equation}\label{eq7}
		p(M_k|d)=\frac{p(M_k) p(d|M_k)}{p(d)}\mbox{ for } k=1,2,...,N. 
	\end{equation} 
	
	The posteriors ratio gives us a measure of the relative plausibility between two models $M_{k}$ and $M_{j}$ as
	\begin{equation}\label{eq8}
		\frac{p(M_k|d)}{p(M_j|d)}=\frac{p(d|M_k)}{p(d|M_j)}\ \frac{p(M_k)}{p(M_j)}  \ ;\ k\neq j.
	\end{equation}
	If we do not have any a priori reason to prefer one model over another before considering the data, $p(M_{k})=p(M_{j})$, so that we can define
	\begin{equation}\label{eq9}
		BF_{kj}=\frac{p(d|M_k)}{p(d|M_j)} \ ;\ k\neq j.
	\end{equation} 
	This expression is called the Bayes factor \citep[see][]{Kass1995} which equals the ratio of marginal likelihoods in Equation (\ref{eq6}) in the case of equal priors.
	
	Once the Bayes factors are computed, the level of evidence for one model against the alternative is assessed by following the criteria for evidence classification developed by \citet{Kass1995}, see Table \ref{tab1}. 

\begin{table}
	\caption{\citet{Kass1995} evidence classification.\label{tab1}}
	\centering
	\begin{tabular}{c c}
		\hline\hline
		$2\ ln BF_{kj}$ & Evidence\\
		\hline
		0-2 & Not Worth more than a bare Mention (NWM)\\
		2-6 & Positive Evidence (PE)\\
		6-10 & Strong Evidence (SE)\\
		$>$10 & Very Strong Evidence (VSE)\\
		\hline
	\end{tabular}
\end{table}

	These general concepts and methods are applicable as long as we first specify priors and likelihood functions. In this work, we have adopted independent priors for model parameters, so that the global prior is given by the product of individual priors
	\begin{equation}\label{eq10}
		p(\mbox{\boldmath$\theta$}|M)=\prod_{i=1}^{n} p(\theta_i|M).
	\end{equation}
	Further, we consider that each parameter lies on a given plausible range, with all values being equally probable a priori. This defines uniform priors of the form
	\begin{equation}\label{eq11}
		p(\theta_i|M)=\frac{1}{\theta_{i}^{\rm max}-\theta_{i}^{\rm min}};\ \theta_i\in(\theta_{i}^{\rm min},\theta_{i}^{\rm max})
	\end{equation}
	and zero otherwise.
	
	As for the likelihood functions, we assume Gaussian profiles, so that the data and the theoretically predicted values are related as
	\begin{equation}\label{eq12}
		p(d|M,\mbox{\boldmath$\theta$})=\frac{1}{\sqrt{2\pi}\sigma_{\Delta}}e^{-\frac{\left(\Delta_{\rm obs}-\Delta_{\rm theor}\right)^2}{2\sigma^2_{\Delta}}},
	\end{equation}
	with $\Delta$ representing the observable and $\sigma_{\Delta}$ its associated uncertainty.
	
	\subsection{Numerical Integration}
	The computation of marginal likelihoods and marginal posteriors using Equations (\ref{eq5}) and (\ref{eq6}) requires the computation of integrals in the parameter space. In low-dimensional parameter spaces, such as the ones considered in this work, the use of direct numerical integration techniques is still feasible from the point of view of computational cost. Nevertheless, we have additionally used Markov Chain Monte Carlo sampling and Monte Carlo integration \citep{Robert2004} to compute marginal posteriors and marginal likelihoods, respectively. The comparison between both types of approaches gives robustness to the obtained results.
	
	The marginal posteriors in Equation (\ref{eq6}) are computed from the Markov Chain Monte Carlo (MCMC) sampling of the global posterior over the space of parameters for each considered model. The marginal likelihoods in Equation (\ref{eq5}) result from Monte Carlo (MC) integration. In this case, MC integration is carried out by the average of all values resulting from the evaluation of the likelihood function, at the points of the parameter space from the MCMC sampling of the normalized prior. These two integration methods are applied employing the \textit{emcee} package of Python \citep{emcee}.
	
	\section{RESULTS}\label{results}
	Before considering the problem of assessing the relative plausibility of the three considered damping mechanisms in explaining the observed time scales, we perform the inference of physical parameters under the assumption that each mechanism is true. Then, our model comparison tools are applied for the computation of the relative plausibility of the three damping models, using possible values for the damping rate and its uncertainty. Finally, these techniques are applied to a large sample of transverse loop oscillations events for which the Bayes factors are computed.

\begin{figure*}
	\figurenum{1}
	\centering
	\includegraphics[trim={0cm 0cm 1.5cm 0cm},scale=0.45]{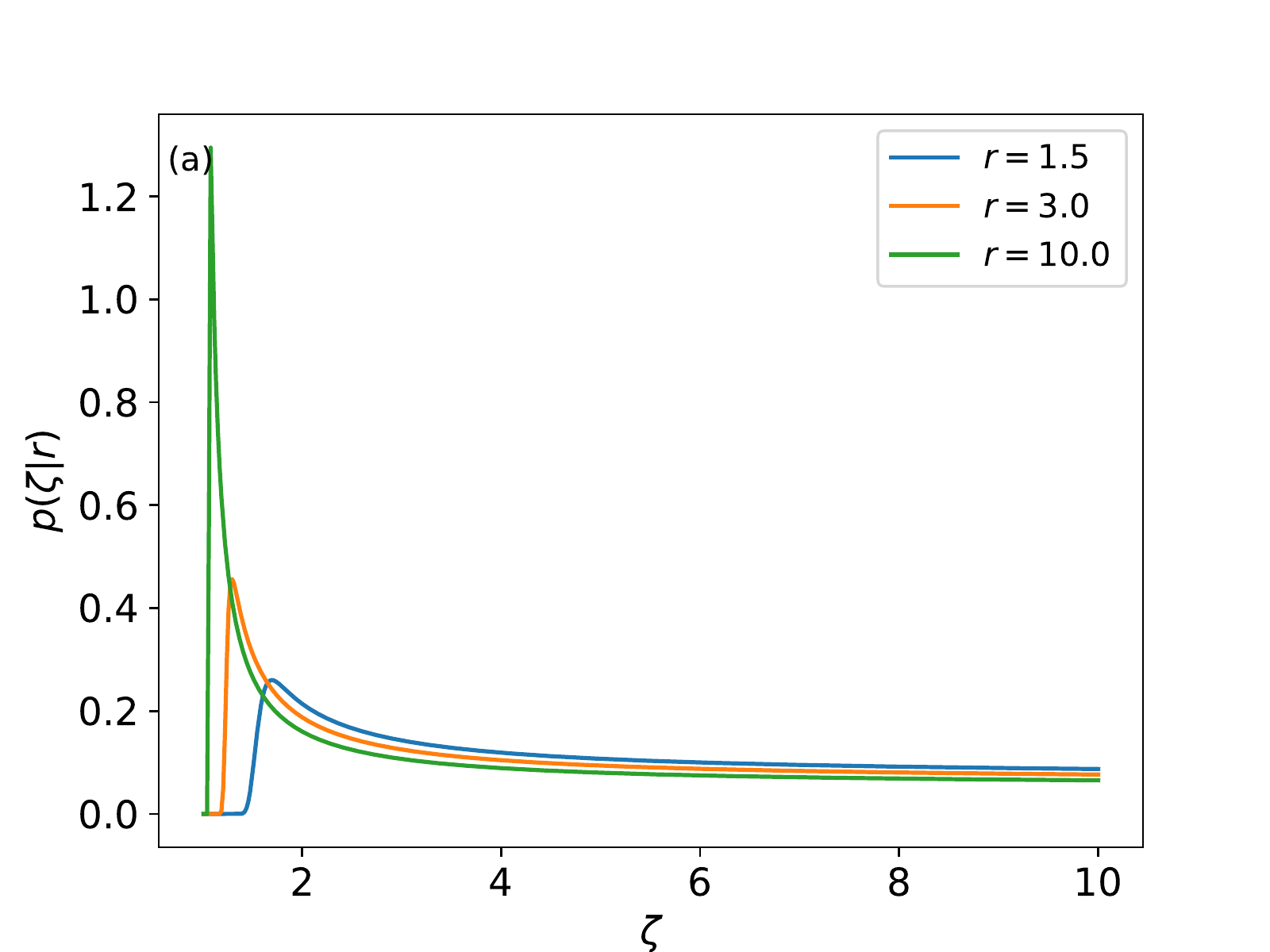} 
	\includegraphics[trim={0cm 0cm 1.5cm 0.cm},scale=0.45]{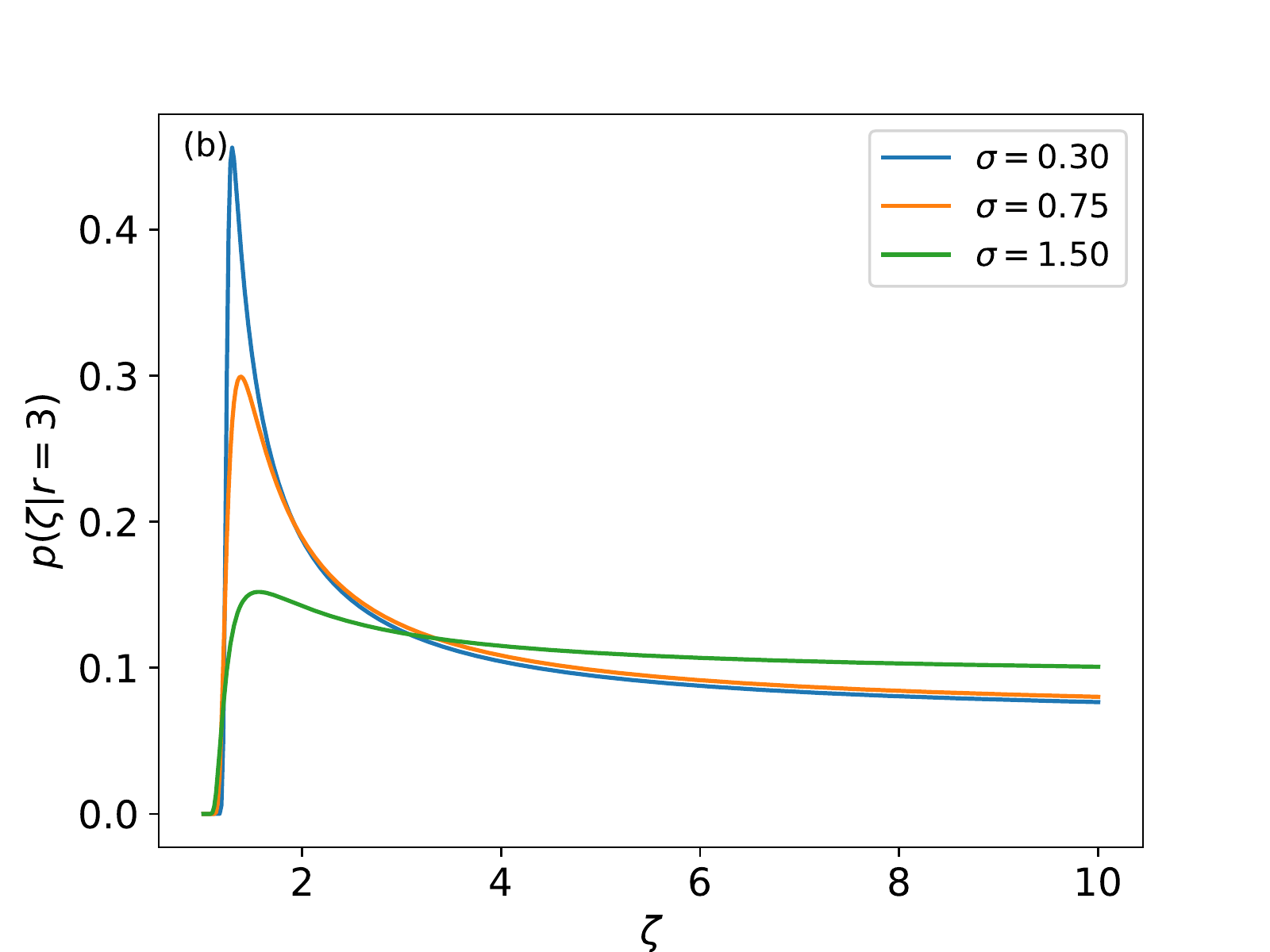}\\
	\includegraphics[trim={0.cm 0cm 1.5cm 0cm},scale=0.45]{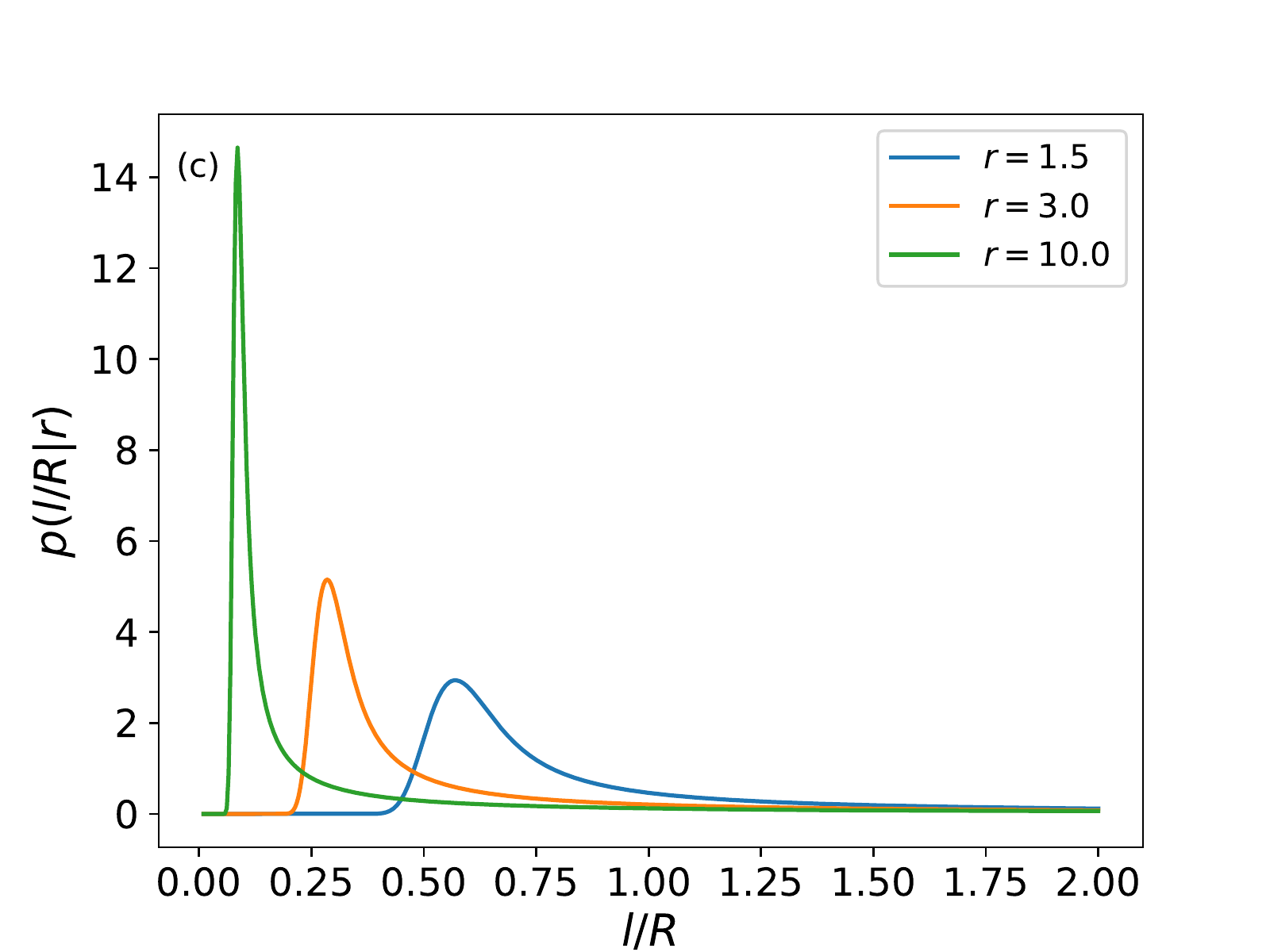}
	\includegraphics[trim={0.cm 0cm 1.5cm 0cm},scale=0.45]{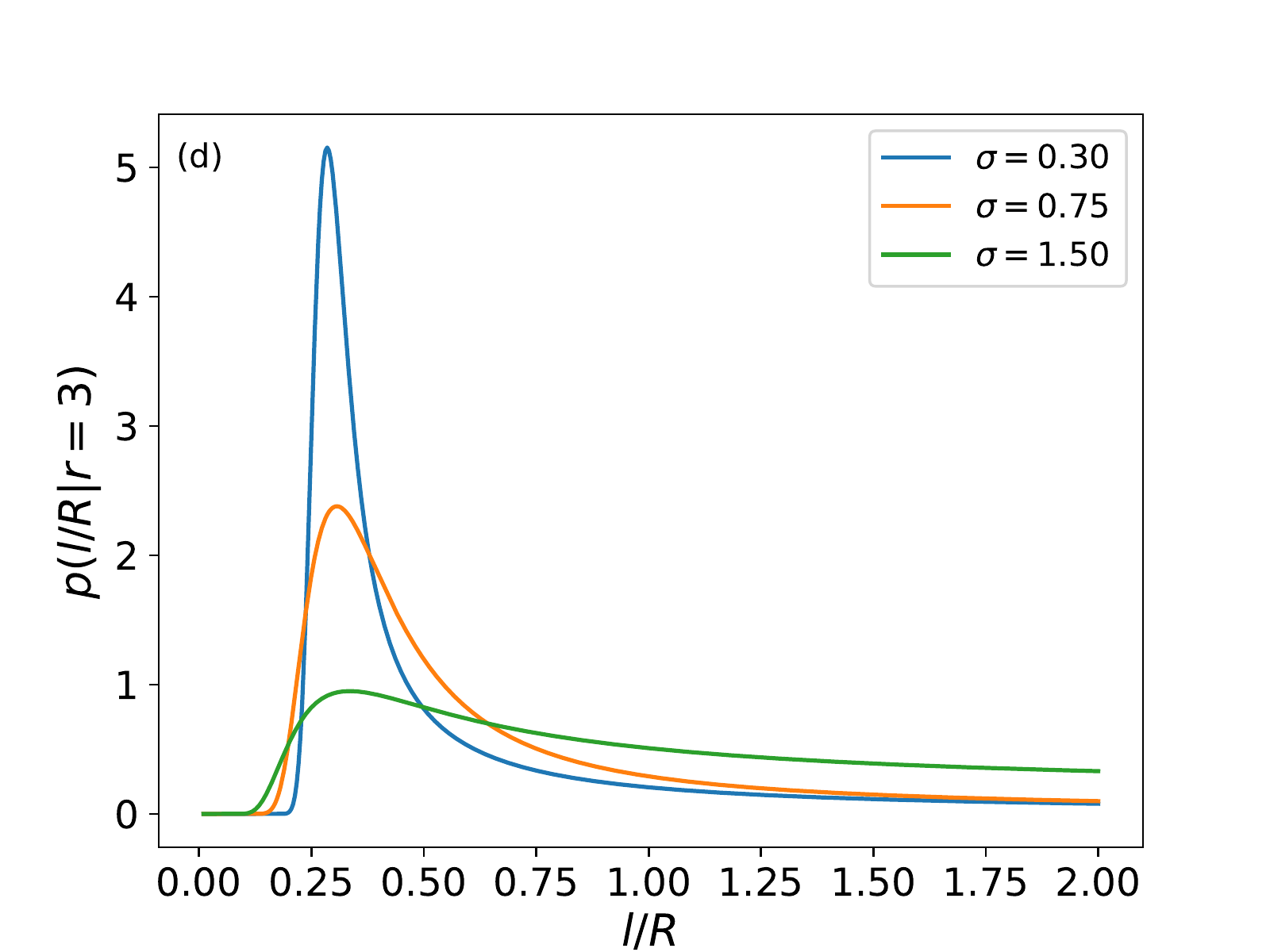}\\
	\includegraphics[trim={0.cm 0cm 0.25cm 0cm},scale=0.45]{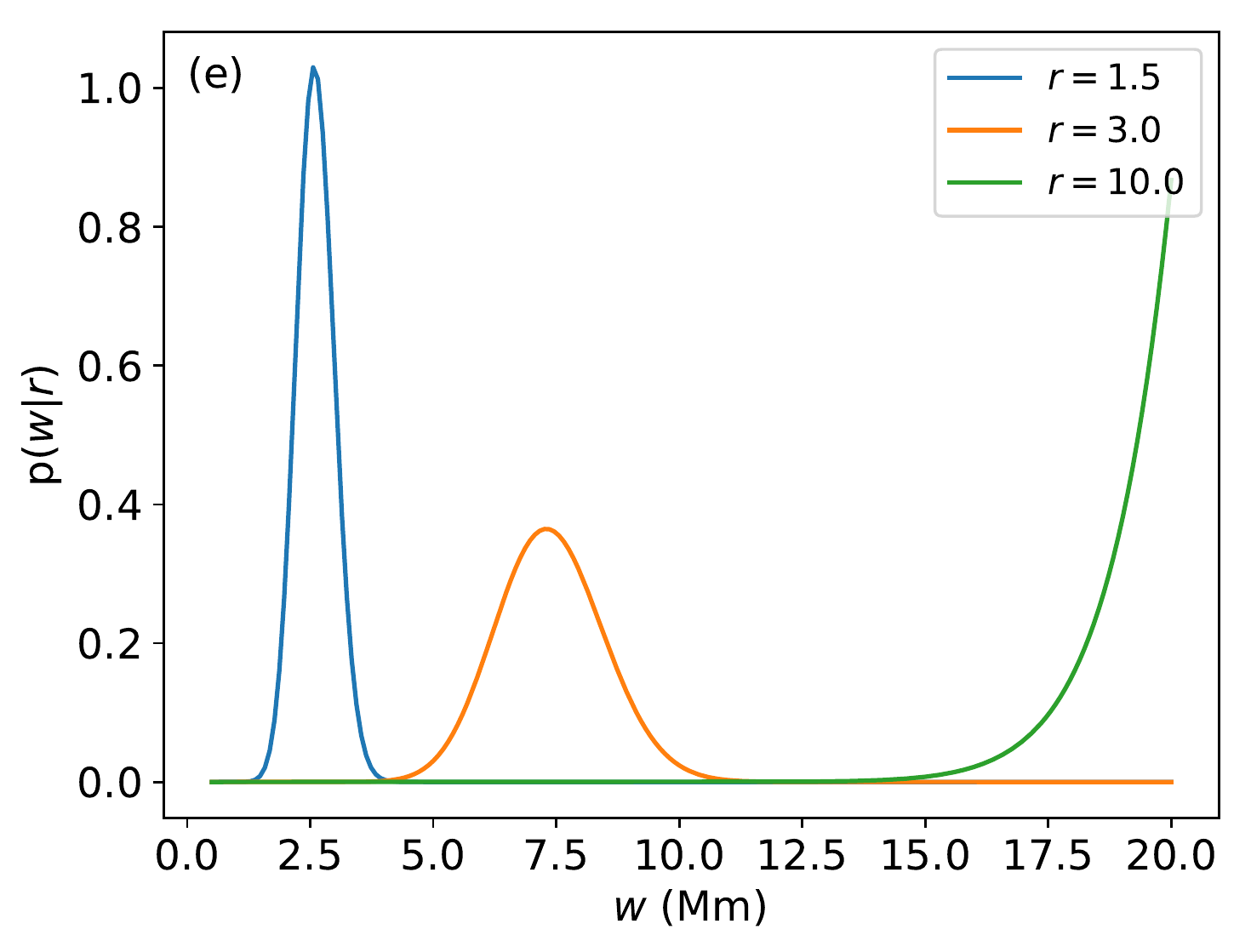}
	\includegraphics[trim={0.cm 0cm 0.25cm 0cm},scale=0.45]{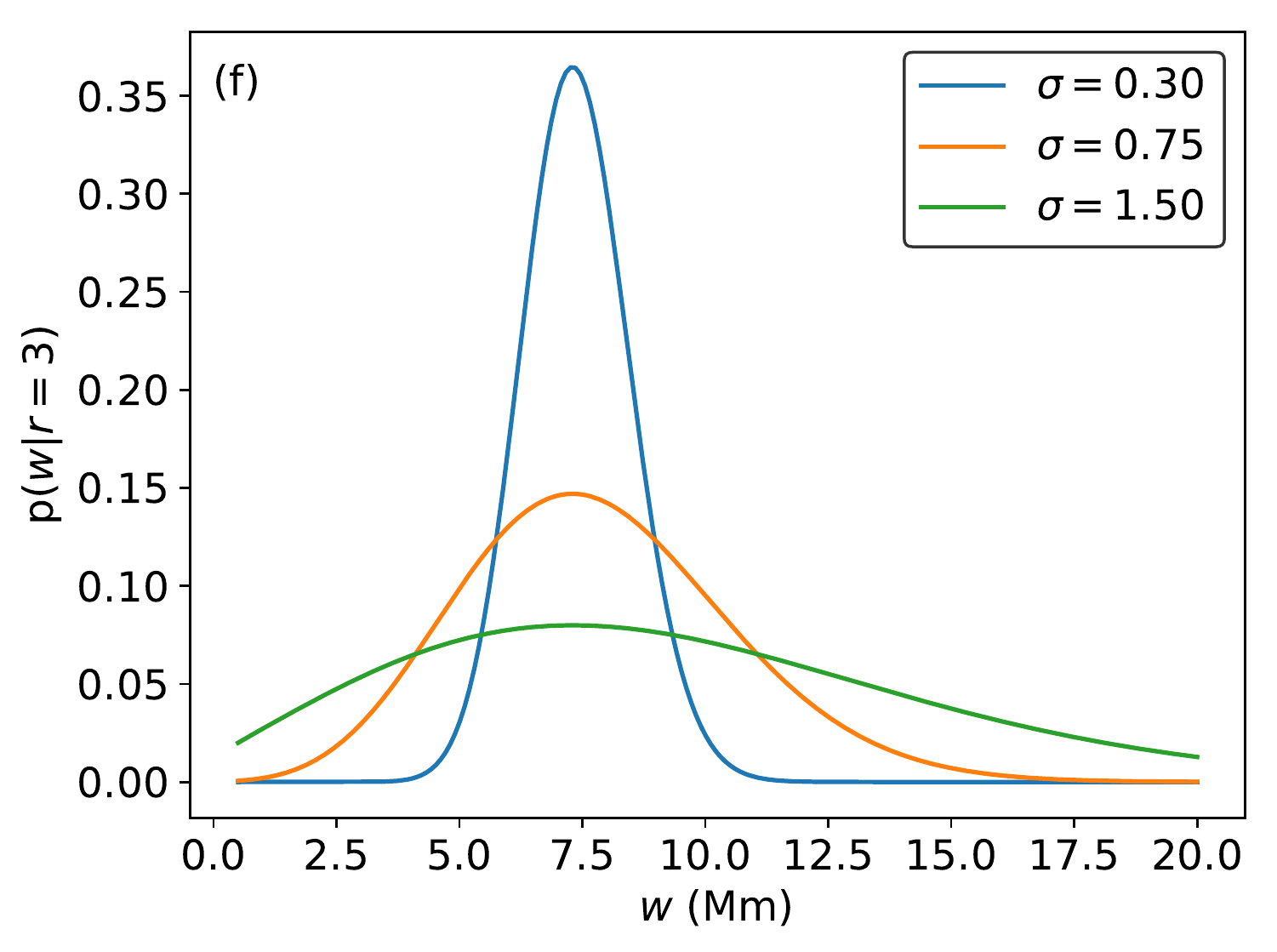}\\
	\includegraphics[trim={0.cm 0.cm 0.25cm 0cm},scale=0.45]{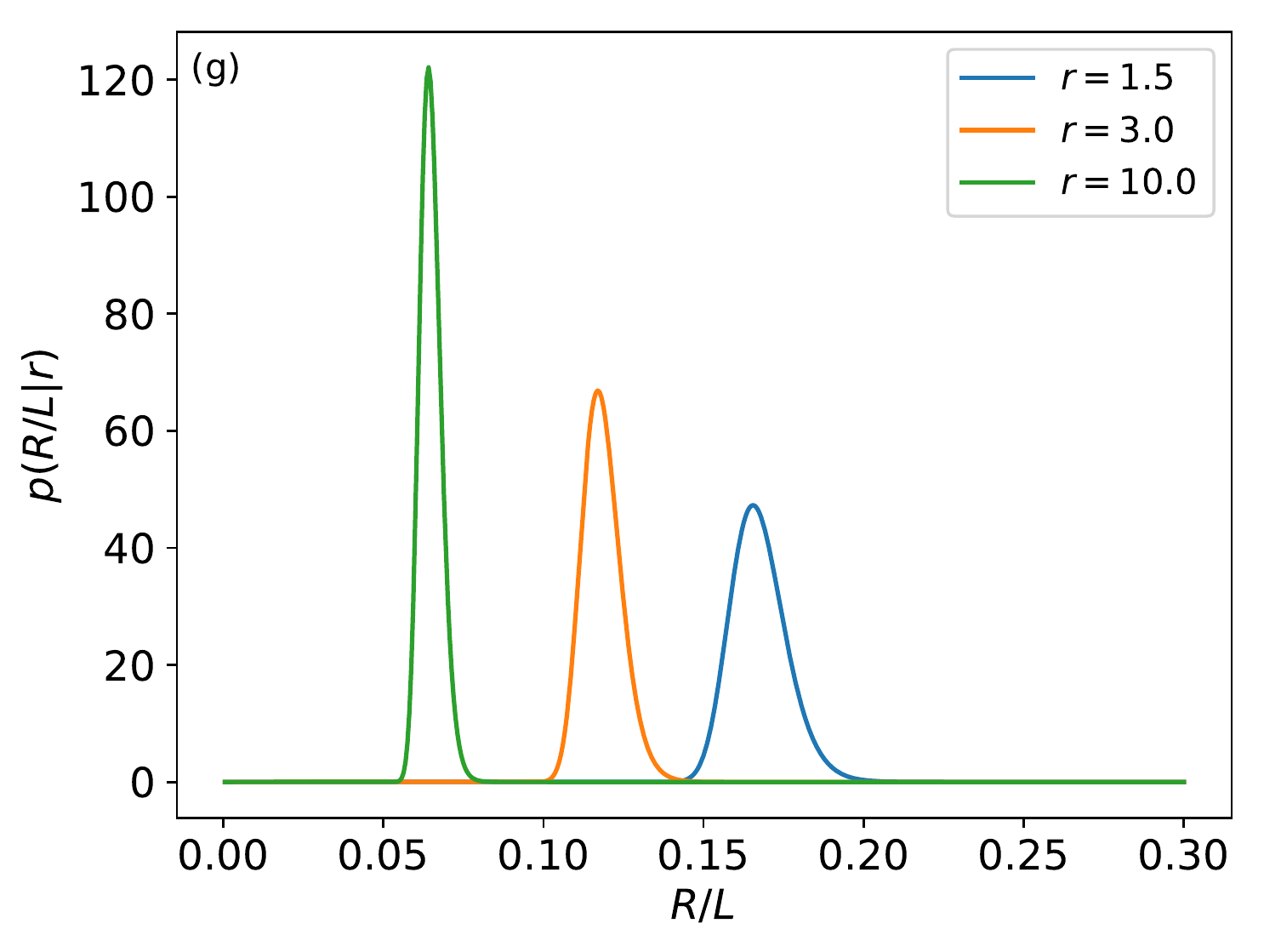}
	\includegraphics[trim={0cm 0.cm 0.25cm 0cm},scale=0.45]{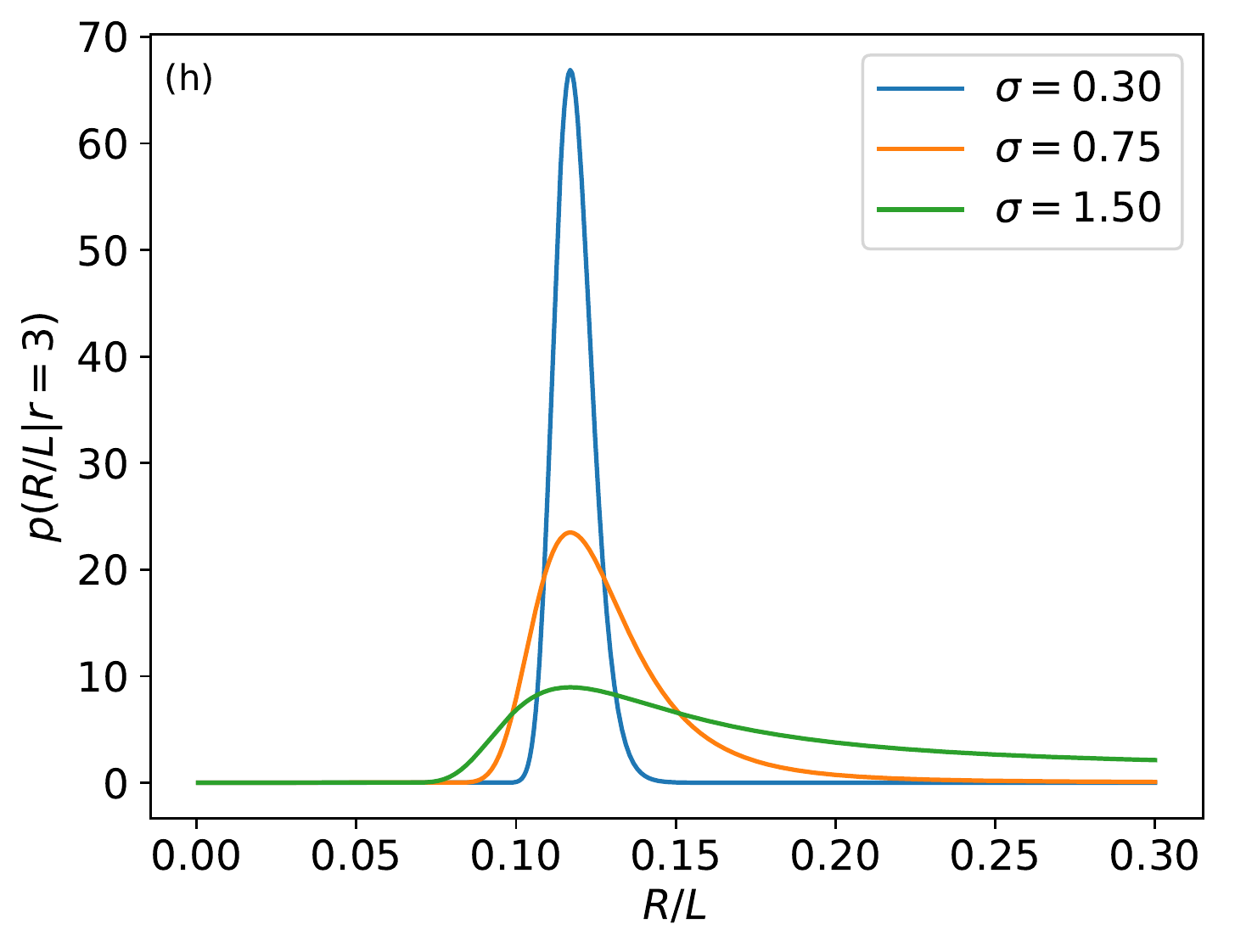}\\
	\caption{Marginal posterior distributions for coronal loop physical parameters under the assumption of resonant damping, (a)-(d); phase-mixing, (e) and (f); and wave leakage, (g) and (h). The left-hand side panels show the results for different values of the damping ratio, $r$, with an uncertainty of 10\%. The right-hand side panels show the results for a fixed damping ratio, $r$=3, and three values for its uncertainty. \label{fig:f1}}	
\end{figure*}

	\subsection{Parameter Inference}\label{inference results}
	For each mechanism, the damping is determined by different coronal loop physical parameters that cannot be directly measured. The theoretical predictions for the observable damping ratio are given by Equations (\ref{eq1}), (\ref{eq2}), and (\ref{eq3}). The inference using observed damping ratios can in principle provide information about the cross-field density inhomogeneity, in the case of resonant absorption and phase mixing, and the coronal loop radius to length ratio in the case of wave leakage. We consider different possible values for the damping ratio and its uncertainty and perform Bayesian inference using Equation (\ref{eq4}), defining for each model Gaussian likelihoods according to Equation (\ref{eq12}) and uniform priors, defined in Equation (\ref{eq11}). Once the full posteriors are known, the marginalization in the corresponding parameter space (Equation \ref{eq6}) leads to the probability density function for each unknown parameter. The resulting marginal posteriors are shown in Figure \ref{fig:f1}. The left hand side panels show results for different values of the observable damping rate with a fixed uncertainty of 10\%. The right hand side panels show results for a fixed observable damping rate and three different values for the uncertainty on the observable damping ratio.

\begin{figure*}
	\figurenum{2}
	\centering
	\includegraphics[scale=0.55]{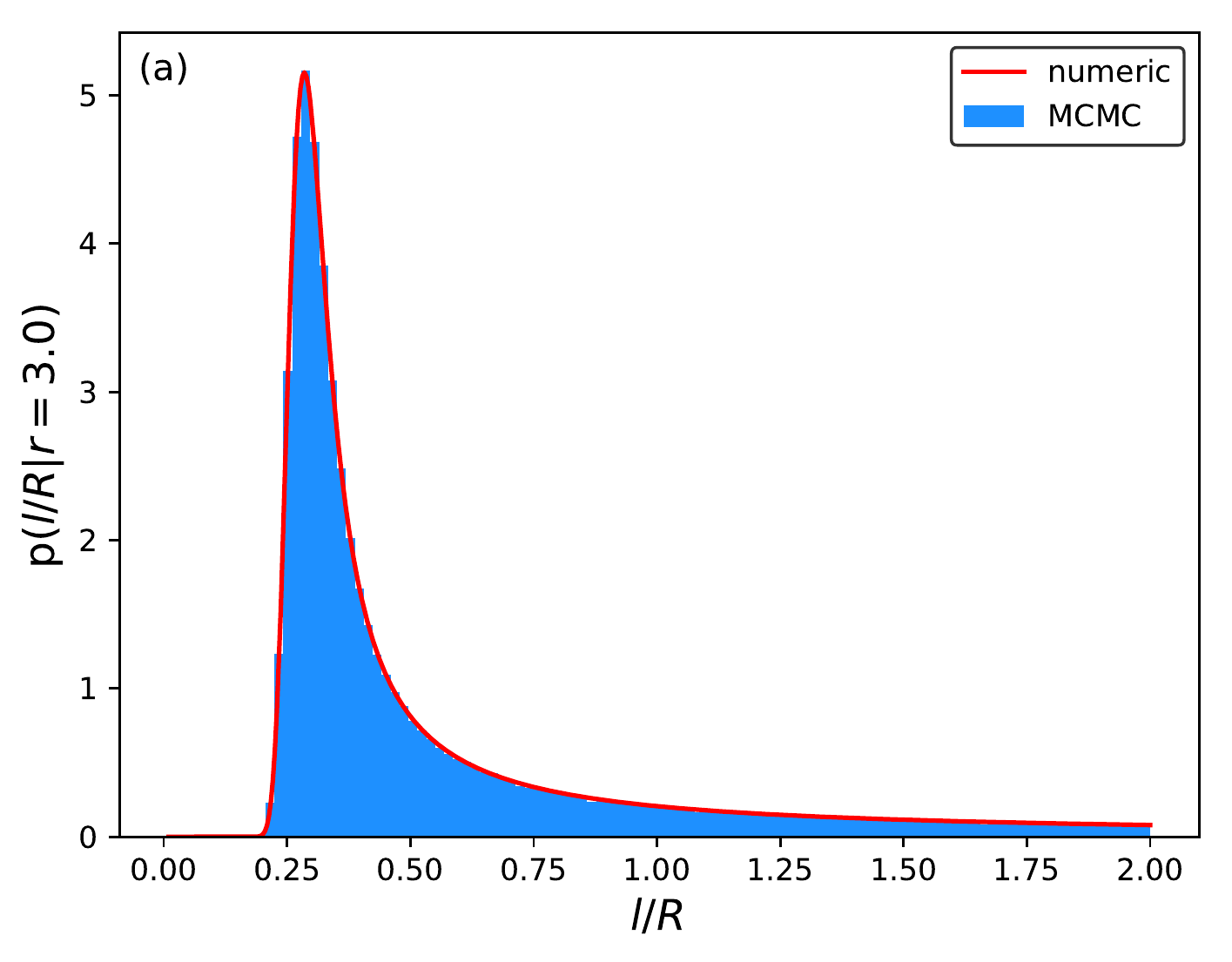} 
	\includegraphics[scale=0.55]{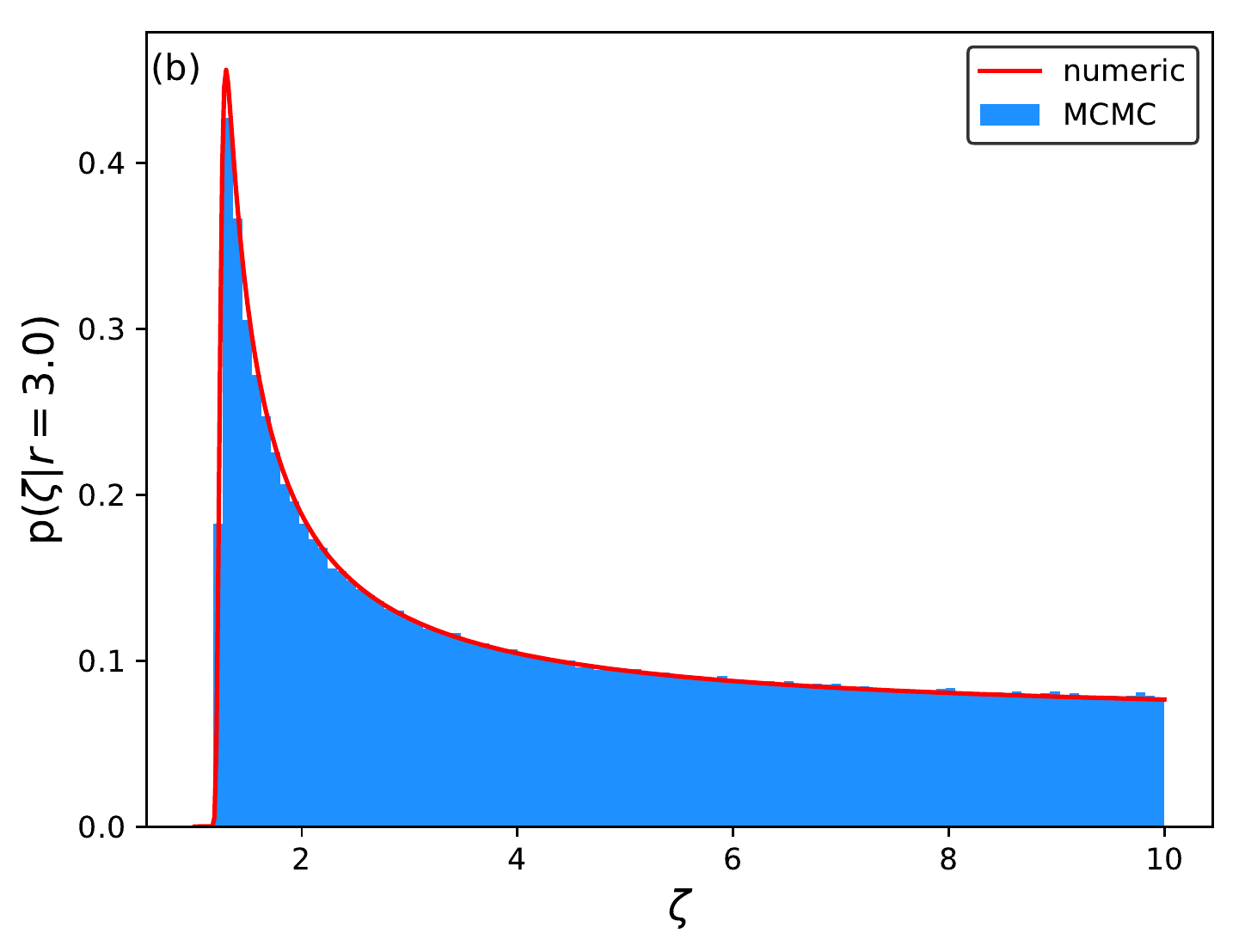}\\
	\includegraphics[scale=0.55]{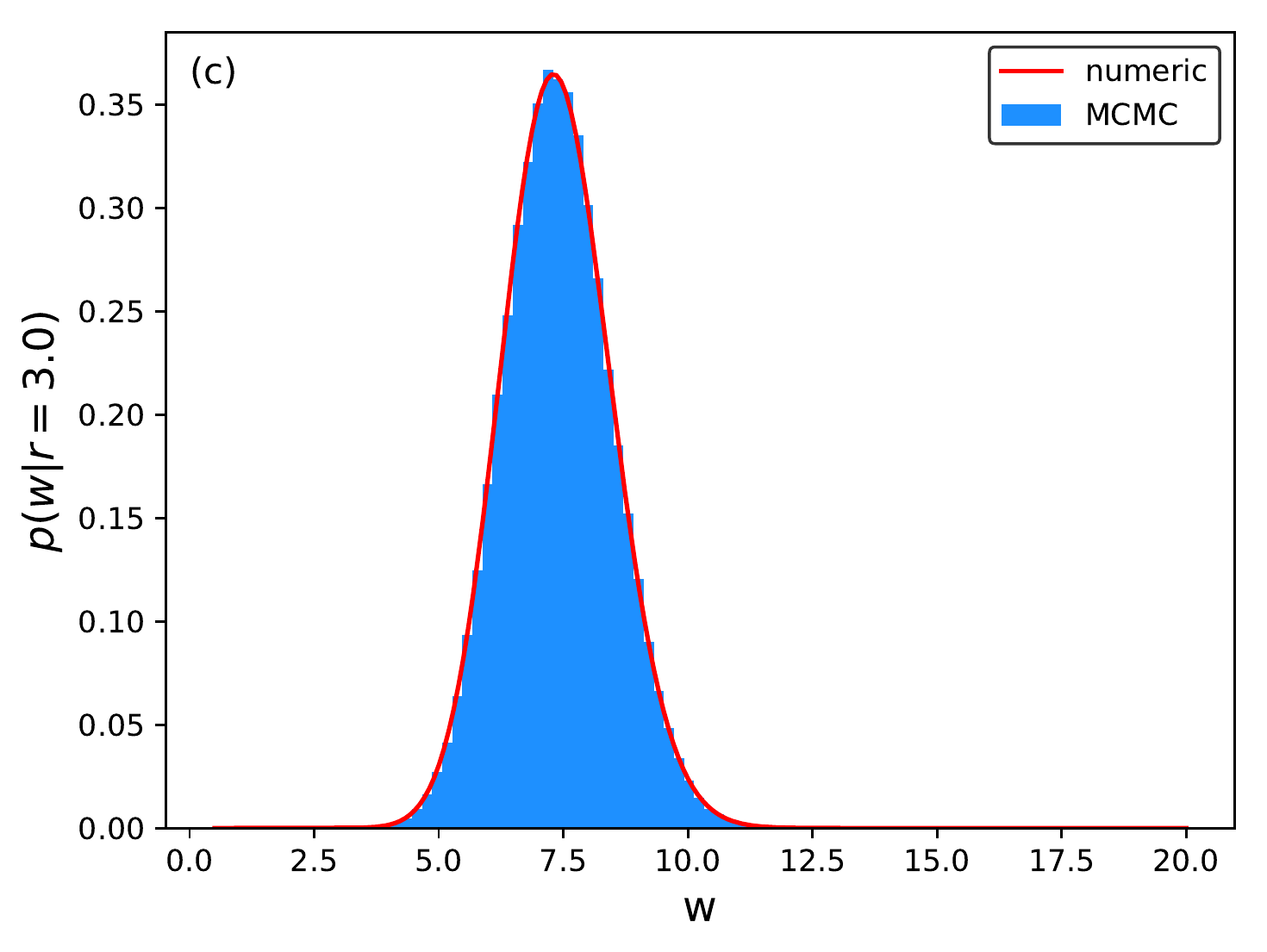}
	\includegraphics[scale=0.55]{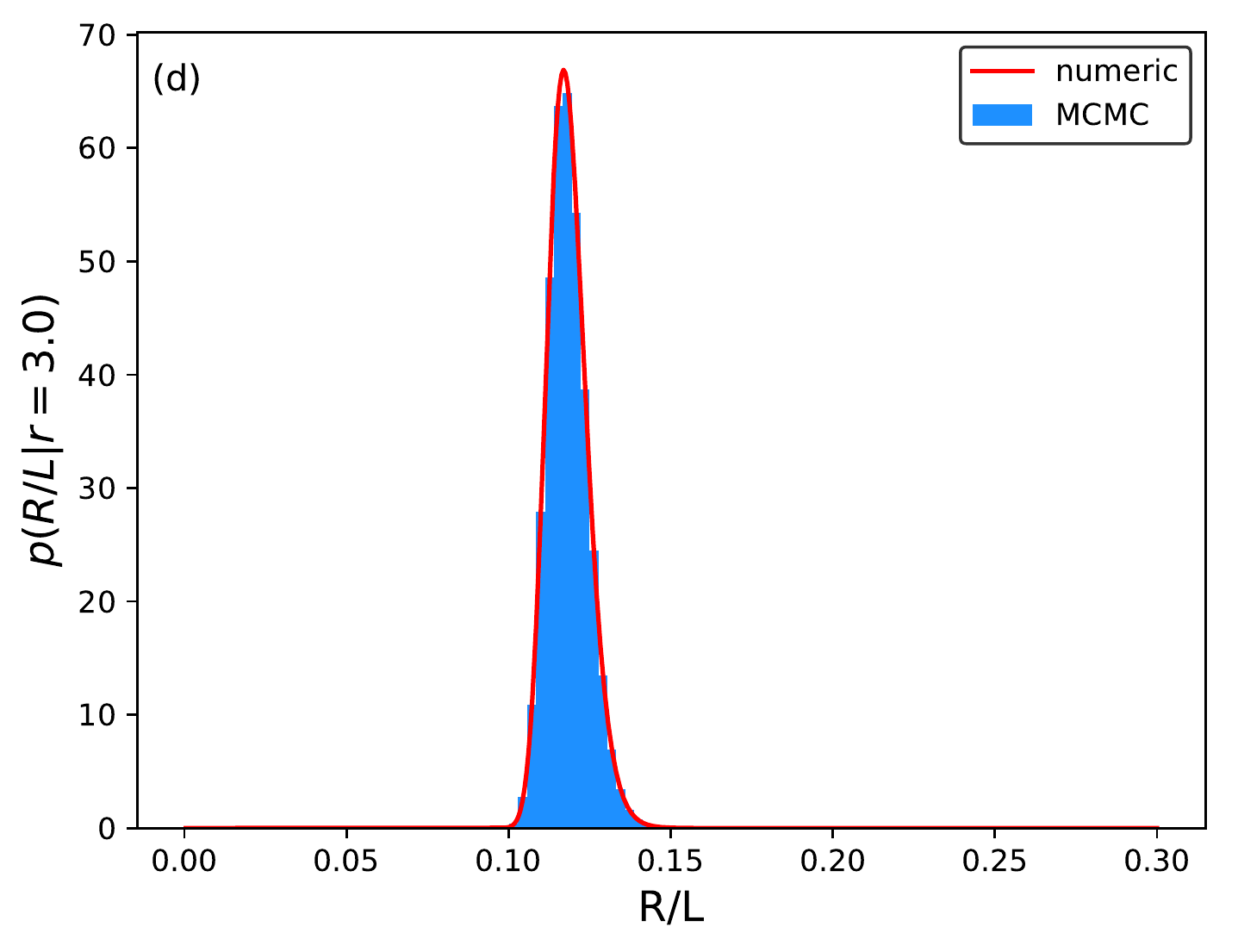}\\		
	\caption{A comparison between the marginal posterior distributions obtained by direct numerical integration (in red) and MCMC sampling (blue histograms) for a case with $r$=3, and an uncertainty of 10\%.\label{fig:f2}}	
\end{figure*}

	Consider first resonant absorption. For this mechanism, such inversion problem was first solved by \citet{Arregui2011} using a MCMC sampling of the posterior by using observed values for periods and damping times. Later on, \citet{Arregui2014} presented solutions akin to those presented here, by making use of the properties of the joint probability of data and parameters. Figures \ref{fig:f1}a-d show the results for this mechanism. For this inference problem,  $\mbox{\boldmath$\theta$}=\{\zeta, l/R\}$ and $d=\tau_{\rm d}/P$.  In principle, both the density contrast, $\zeta$, and the transverse inhomogeneity length scale, $l/R$, can be properly inferred, although as already noted by \citet{Arregui2014} the posteriors for the density contrast show long tails. The posterior densities peak at lower values for both unknowns the larger the damping ratio is (see Figures \ref{fig:f1}a and \ref{fig:f1}c) and their distributions have broader shapes the larger the uncertainty in the observed damping ratio (see Figures \ref{fig:f1}b and \ref{fig:f1}d). These results indicate that, in principle, the assumption that resonant absorption is the true mechanism that produces the damping of coronal loop oscillations together with measurements of the damping ratio enables us to infer the two parameters that control the cross-field density inhomogeneity.

\begin{table*}
	\caption{Inferred parameters from Figure \ref{fig:f2}.\label{tab2}}	
	\centering
	\begin{tabular}{c c c c c}
		\hline\hline
		Method& $\zeta$& $l/R$& $w$& $R/L$\\
		&	&	& (Mm)& 	\\
		\hline
		Numeric&$4.1^{+3.9}_{-2.4}$&$0.4^{+0.5}_{-0.2}$&$7.5^{+1.1}_{-1.1}$&$0.12^{+0.01}_{-0.01}$\\
		MCMC&$4.1^{+3.8}_{-2.5}$&$0.4^{+0.5}_{-0.1}$&$7.4^{+1.1}_{-1.0}$&$0.12^{+0.01}_{-0.01}$\\
		\hline\hline
	\end{tabular}
\end{table*}
	
	For phase mixing, the theoretical prediction in Equation (\ref{eq2}) does not enable to factorize the damping rate as a sole function of model parameters since the oscillation period is present in the right-hand side of this equation. For this reason, we perform the inference by assuming particular values for the period in the range $P\in[150,1250]$s. This leaves us with an inversion problem with one observable, $d=\tau_{d}/P$, and one unknown, $\mbox{\boldmath$\theta$}=\{w\}$. Figures \ref{fig:f1}e and \ref{fig:f1}f show the results for this mechanism. The posteriors computed for the case P=150 s are only shown to not deface the plot. The posteriors shift toward larger values of $w$ for longer periods. The inference results for this mechanism show that the length scale $w$ can be properly inferred although a larger upper limit in the parameter range is needed for the case r=10. The posteriors shift towards larger values of $w$ the longer the damping ratio is (see Figure \ref{fig:f1}e) and show broader distributions the larger the uncertainty in the observed damping ratio (see Figure \ref{fig:f1}f). These results indicate that the assumption that phase mixing is the true damping mechanism of coronal loop oscillations together with measurements of the damping ratio enables us to infer $w$.

	Regarding wave leakage, the theoretical damping rate in Equation (\ref{eq3}) is only a function of the coronal loop radius to length ratio, $R/L$. This enables us to solve the inference with one observable, $d=\tau_{d}/P$, and one parameter, 
$\mbox{\boldmath$\theta$}=\{R/L\}$. Figures \ref{fig:f1}g and \ref{fig:f1}h show the results for this mechanism. They show that the ratio between the radius and the length of coronal loops can be properly inferred. The posterior distributions shift toward smaller values of $R/L$ the larger the damping ratio is (see Figure \ref{fig:f1}g) and have broader shapes the larger the uncertainty in the observed damping ratio (see Figure \ref{fig:f1}h). These results indicate that assuming that wave leakage is the true mechanism that causes the damping of coronal loop oscillations together with measurements of the damping ratio, a relation among two structural features of coronal loops can be inferred.
	
	Comparing these results obtained from direct numeric integration with the results from MCMC sampling, we can see the  degree of applicability of each method to the inference problem. An example of this comparison is show in Figure \ref{fig:f2}, and the corresponding median values of damping model parameters within a credible interval of 68\% are presented in Table \ref{tab2}. The agreement between the posterior distributions and the minimal differences between the mean values, indicate the robustness of results and the applicability of both direct numerical integration and MCMC to solve this type of inference problems.

	\subsection{Model Comparison}
	In this section, we show the results from the application of Bayesian model comparison to the three considered damping mechanisms. The comparison is performed in two principal ways. First, we consider plausible values for the observed damping rate and compute the marginal likelihoods according to Equation (\ref{eq5}), and the Bayes factors given by Equation (\ref{eq9}) for each model as a function of this observable. Then, the Bayes factors between the damping models are computed as a function of periods and damping times on a wide range of values and for a given observational uncertainty. Finally, the method is applied to a large sample of transverse loop oscillation data to determine in how many of the real events evidence supporting either of the considered mechanisms is found.
		
\begin{figure*}
	\centering
	\figurenum{3}
	\includegraphics[scale=0.5]{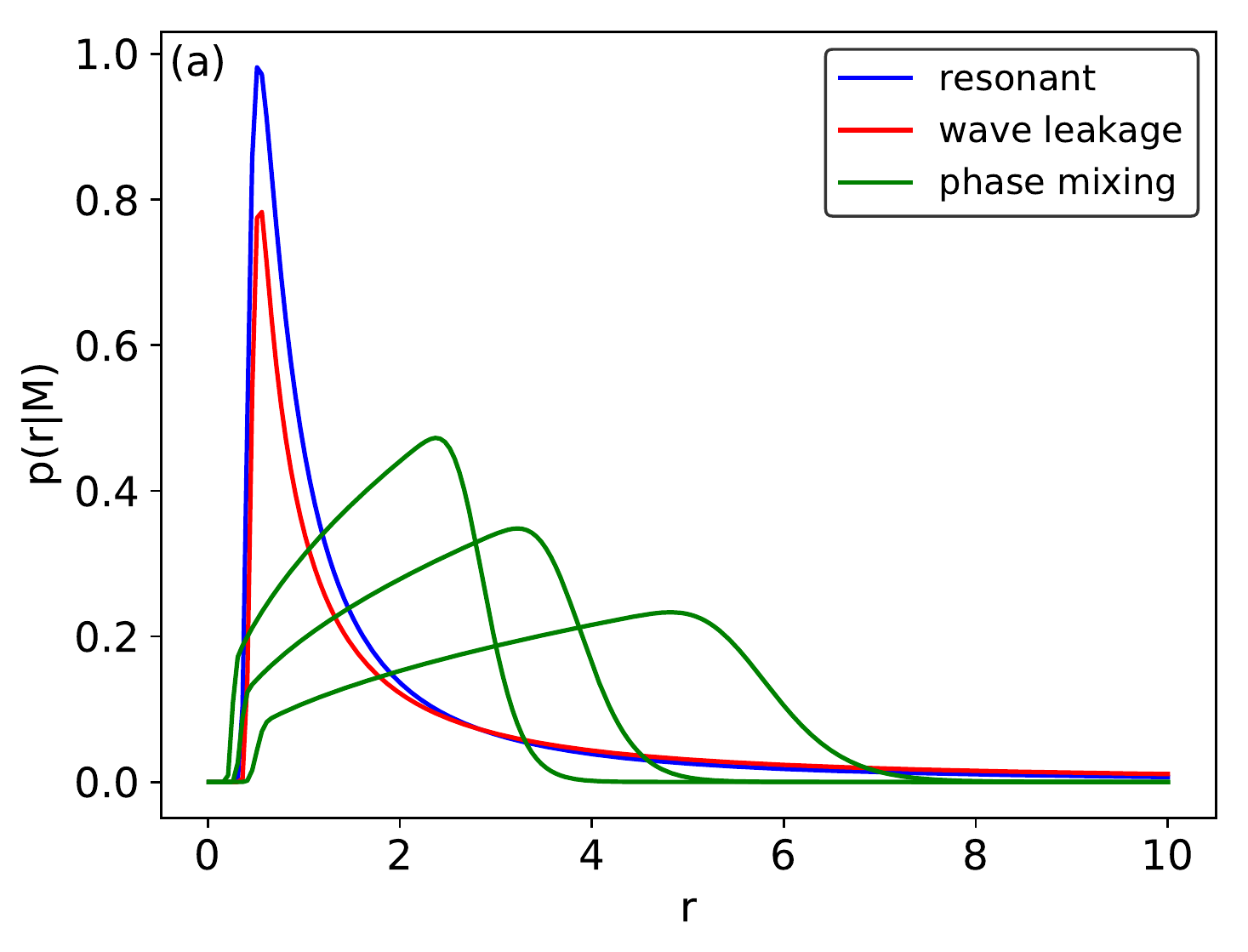}
	\includegraphics[scale=0.5]{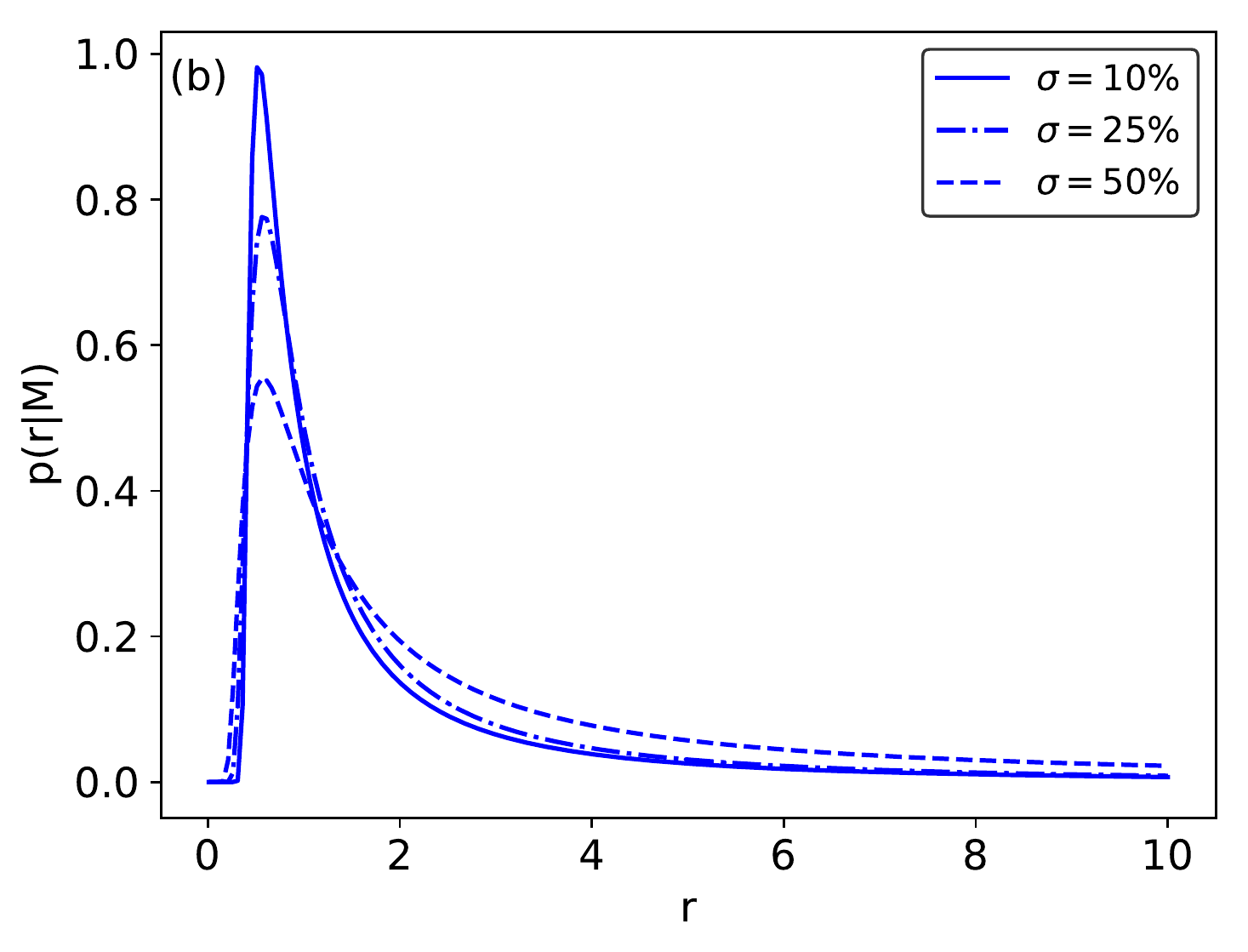}\\
	\includegraphics[scale=0.5]{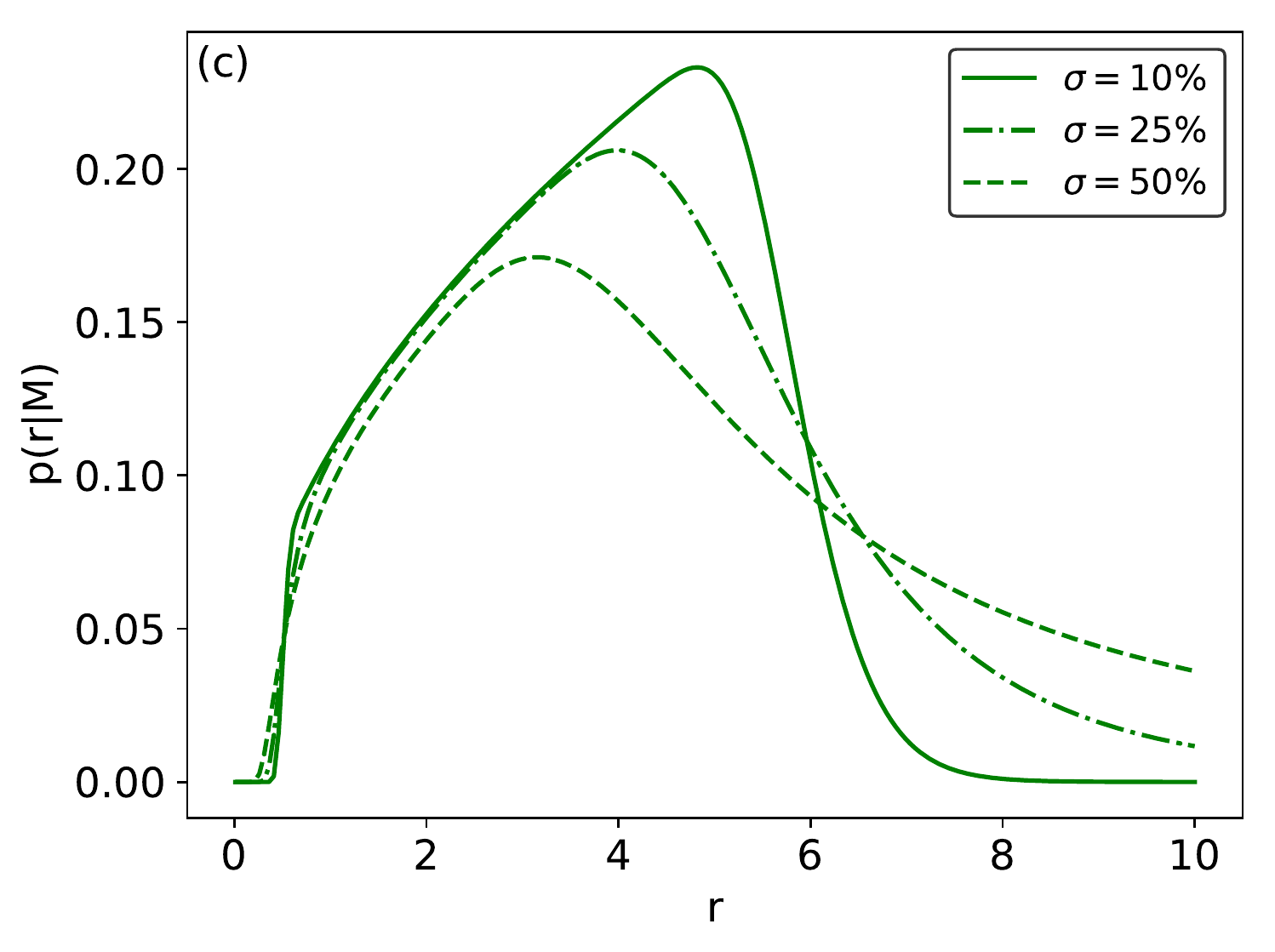}
	\includegraphics[scale=0.5]{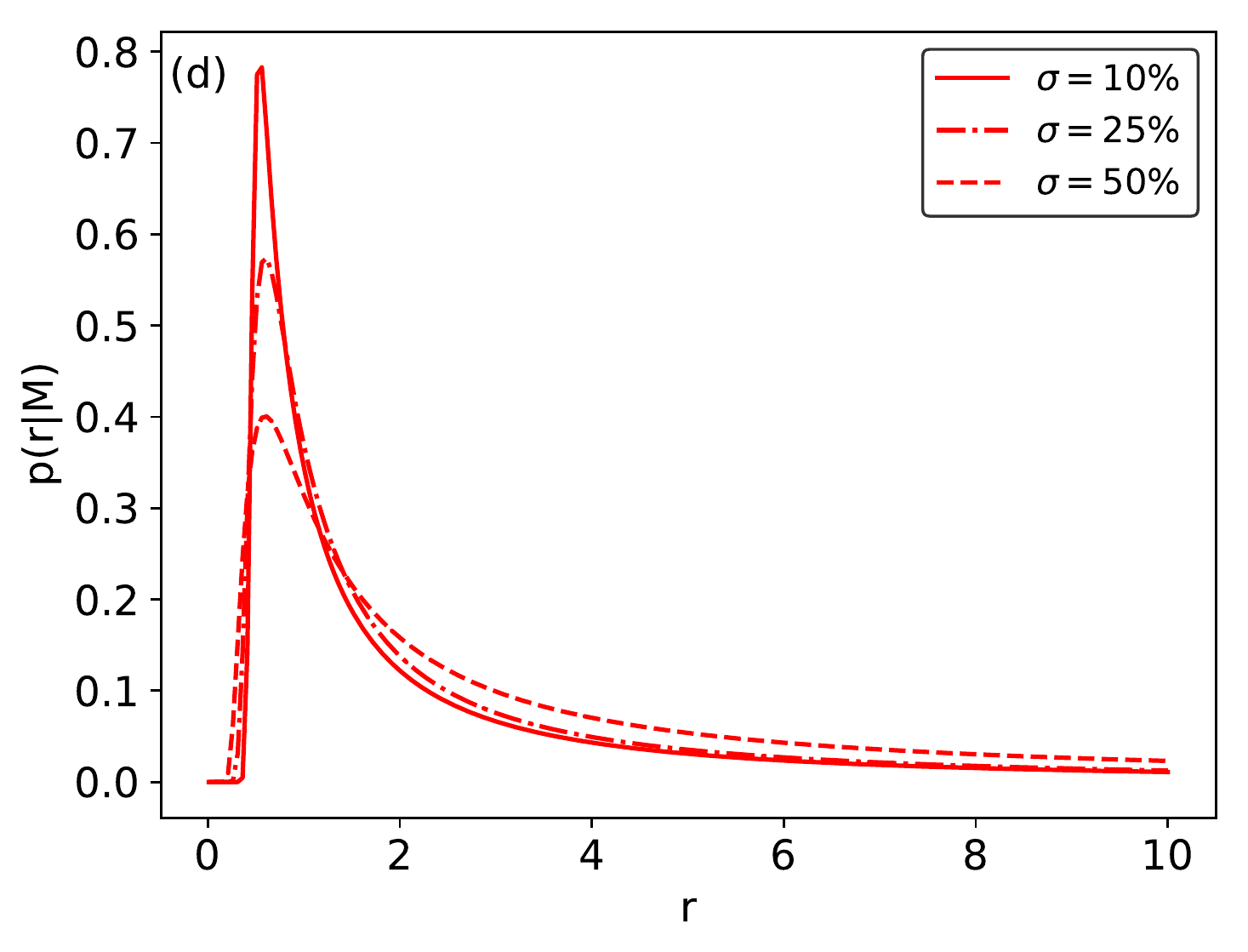}\\
	\caption{(a) Marginal likelihoods computed using Equation (\ref{eq5}) for the three considered models as a function of the observable damping ratio for a fixed value of $\sigma=10\%$. For phase mixing, three fixed values of period ($150,500, \mbox{ and } 1250\ s$ from right to left) have been taken. (b)-(d): Marginal likelihoods for each mechanism separately for three different values of $\sigma=10\%, 25\%, 50\%$. For each mechanism the same color as in panel (a) is applied.\label{fig:f3}}
\end{figure*}

\begin{figure}
	\figurenum{4}
	\centering
	\includegraphics[scale=0.5]{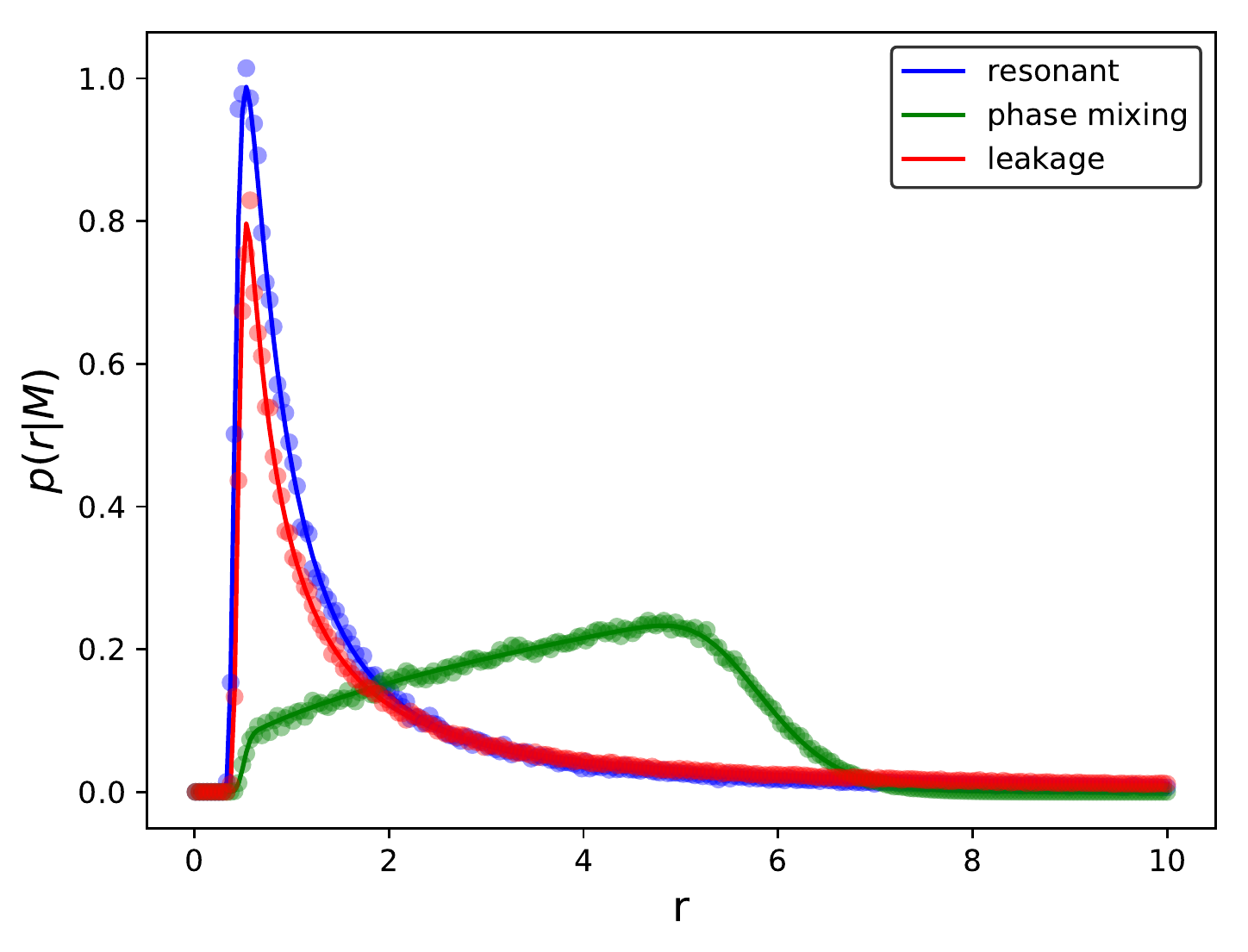}
	\caption{A comparison between the marginal likelihoods obtained by direct numerical integration (continuous lines) and the Monte Carlo integration (filled circles) for the three damping mechanisms as a function of the observable damping ratio with uncertainty of 10\%. For phase mixing, $P=150$ s is fixed. \label{fig:f4}}
\end{figure}

	Figure \ref{fig:f3}a shows the results from the computation of marginal likelihoods for the three considered damping mechanisms, as a function of the damping ratio. Resonant absorption and wave leakage offer more plausible explanations for low damping rate values, while the evidence for phase mixing attains larger values from low to intermediate values of the damping rate. We note that for phase mixing, as the analytic expression for the damping ratio is a function of the wave period, its marginal likelihood are computed for particular values of the period. These results are determined for a fixed value of the uncertainty on the observed damping ratio. Figures \ref{fig:f3}b-d show the marginal likelihoods, separately for each mechanism, for three different values for the error on the measured damping ratio. In all three cases, increasing the uncertainty produces a decrease on the maximum value of the marginal likelihood, as well as some spread out of the distributions. The discussed marginal likelihoods are computed by direct numerical integration, using Equation (\ref{eq5}). Just as with the inference in the previous section, we check that they can equally well be computed by means of Monte Carlo integration. For each damping ratio, we carry out the MCMC sampling of the global prior given by Equation (\ref{eq10}), just as we sample the posterior for the inference. Then, we evaluate the likelihood function in Equation (\ref{eq12}) over the sampled parameters, and finally, we average the results to obtain the marginal likelihood. This is shown in Figure \ref{fig:f4}, where a comparison between both methods to evaluate marginal likelihoods is presented. On the one hand, this result confirms the goodness of direct numerical integration. On the other, it demonstrates that the developed Monte Carlo integration method can be confidently used when the first approach will not be feasible, for example, in cases with a larger dimension of the parameter space.

\begin{figure*}
	\figurenum{5}
	\centering
	\includegraphics[trim={0.2cm 0cm 0.2cm 0cm},scale=0.4]{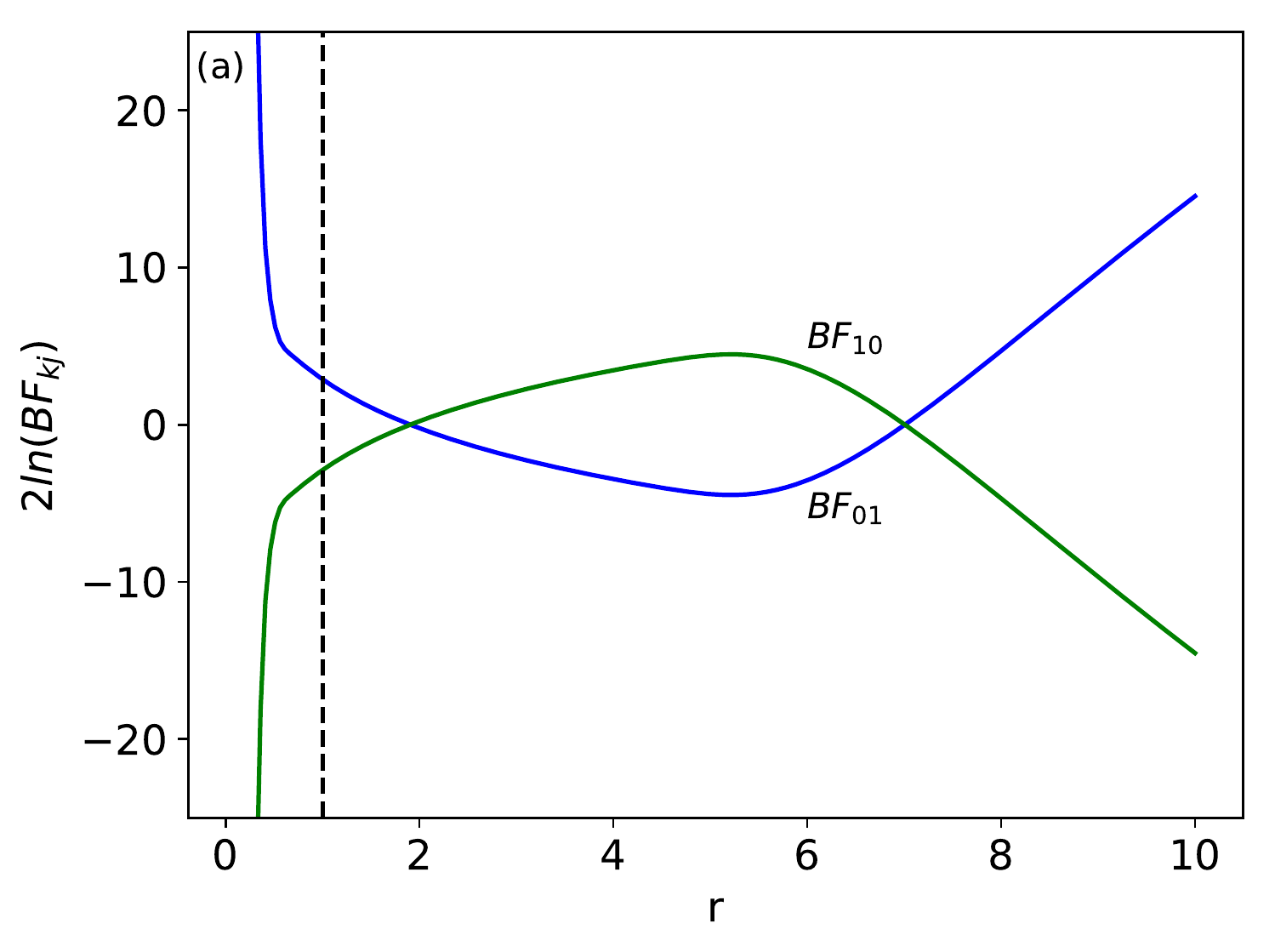}
	\includegraphics[trim={0.2cm 0cm 0.2cm 0cm},scale=0.4]{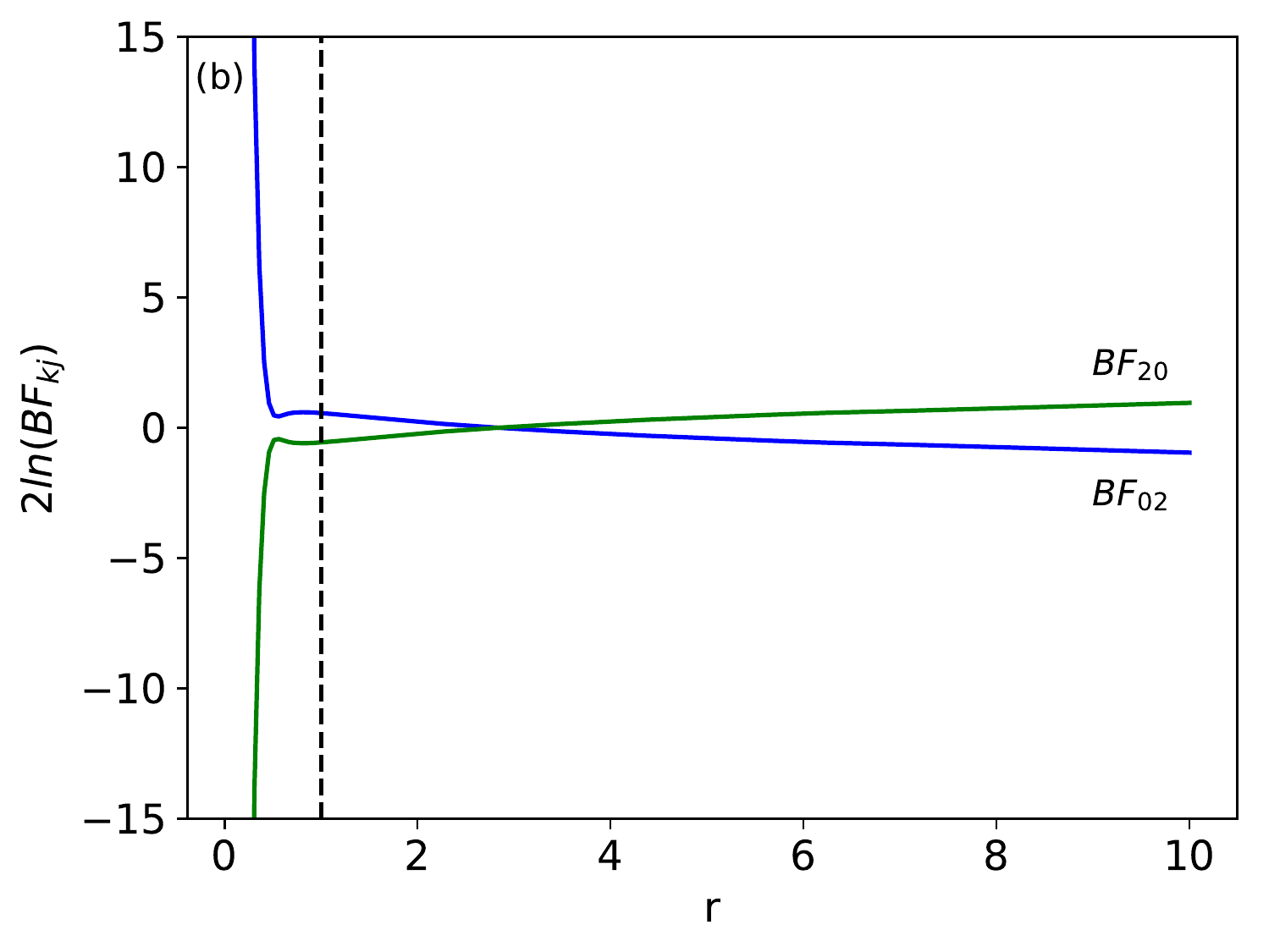}
	\includegraphics[trim={0.2cm 0cm 0.2cm 0cm},scale=0.4]{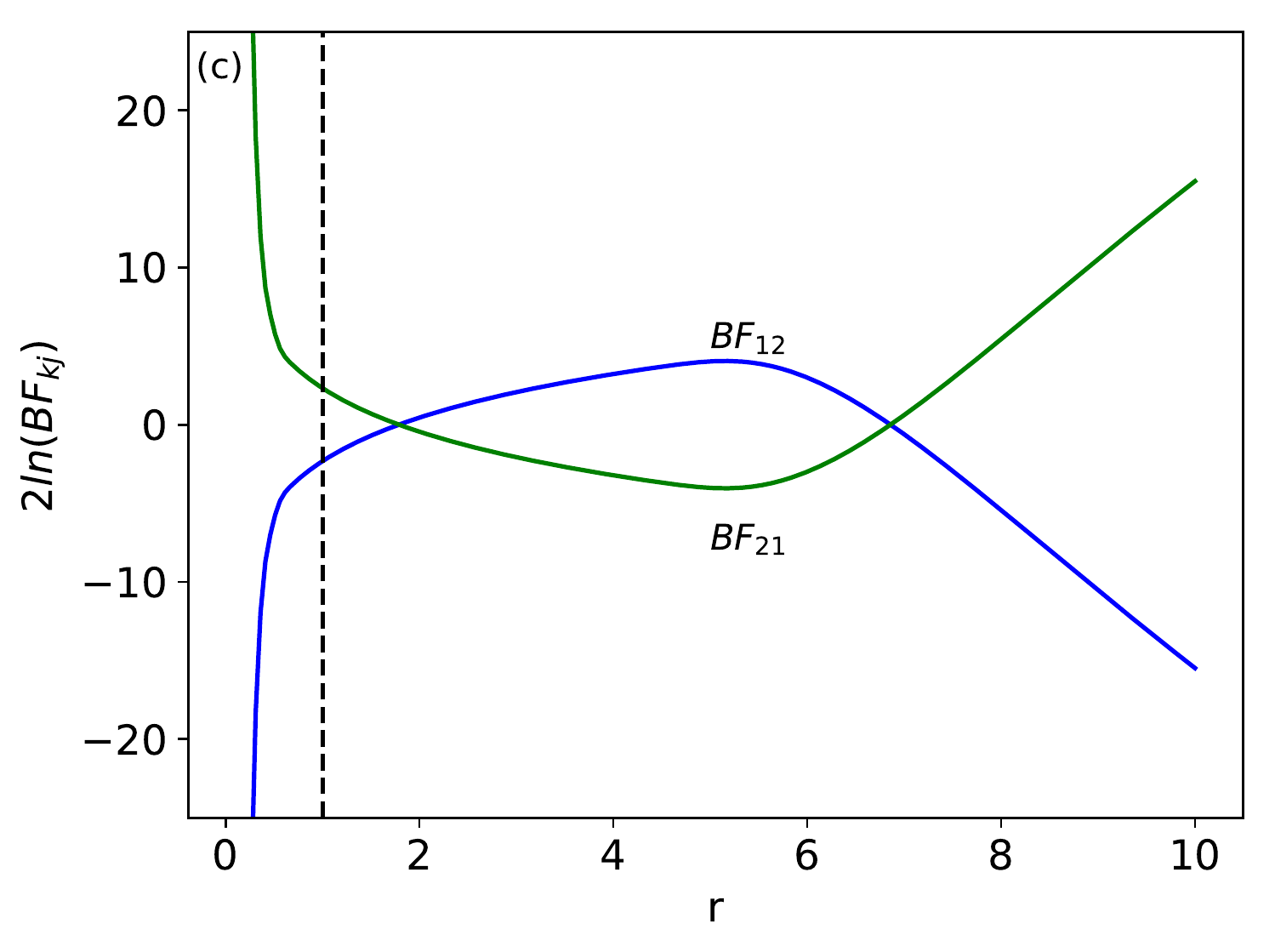}
	\caption{Bayes factors for the one-to-one comparison between resonant absorption, phase mixing, and wave leakage mechanisms, here represented with the subscripts 0, 1, and 2 respectively, in the range $r\in[0.01,10]$ with uncertainty of 10\%. For phase mixing, $P$=150 s is fixed. \label{fig:f5}}
\end{figure*}

	The quantitative assessment of the relative performance between damping mechanisms is given by the computation of the Bayes factors in the one-to-one comparison between the three damping mechanisms. In the following, we use the subscripts 0, 1, and 2, to identify resonant absorption, phase mixing, and wave leakage, respectively. Figure \ref{fig:f5} shows the distribution of Bayes factors as a function of the damping ratio corresponding to one-to-one comparisons between the three mechanisms. An error measurement on the damping ratio of 10\% is assumed. In Figure \ref{fig:f5}a, we see that very strong evidence for resonant absorption in front of phase mixing is obtained for the lowest and the largest considered values of the damping ratio, in between 0 and 0.5 and for r $>$ 9. The first interval corresponds to extremely strong damping regimes in which the damping time is shorter than the oscillation period. The second to rather weak damping regimes. For intermediate damping ratio values, the evidence favors phase mixing, with values for $2lnBF_{10}$ in between 2 and 5. These Bayes factor values correspond to positive evidence for phase mixing in front of resonant absorption, according to the levels of evidence in Table \ref{tab1}. For the comparison between resonant absorption and wave leakage, Figure \ref{fig:f5}b shows very strong evidence in favor of the resonant absorption again for the lowest values of the damping ratio, followed by a decrease of $BF_{02}$ until both Bayes factors intersect. For intermediate and large values of the damping ratio, the evidence is inconclusive since the magnitude of both Bayes factors is below 2. Finally, Figure \ref{fig:f5}c shows the confrontation between phase mixing and wave leakage, with results that are akin to those between resonant absorption and phase mixing in Figure \ref{fig:f5}a. Wave leakage is the favored model for strong and weak damping regimes, while positive evidence for phase mixing is obtained in the intermediate damping regime, with $r\in~(2-7)$.

\begin{figure*}
	\figurenum{6}
	\vspace{0.2cm}
	\includegraphics[trim={1.25cm 0cm 0cm 1cm},scale=0.44]{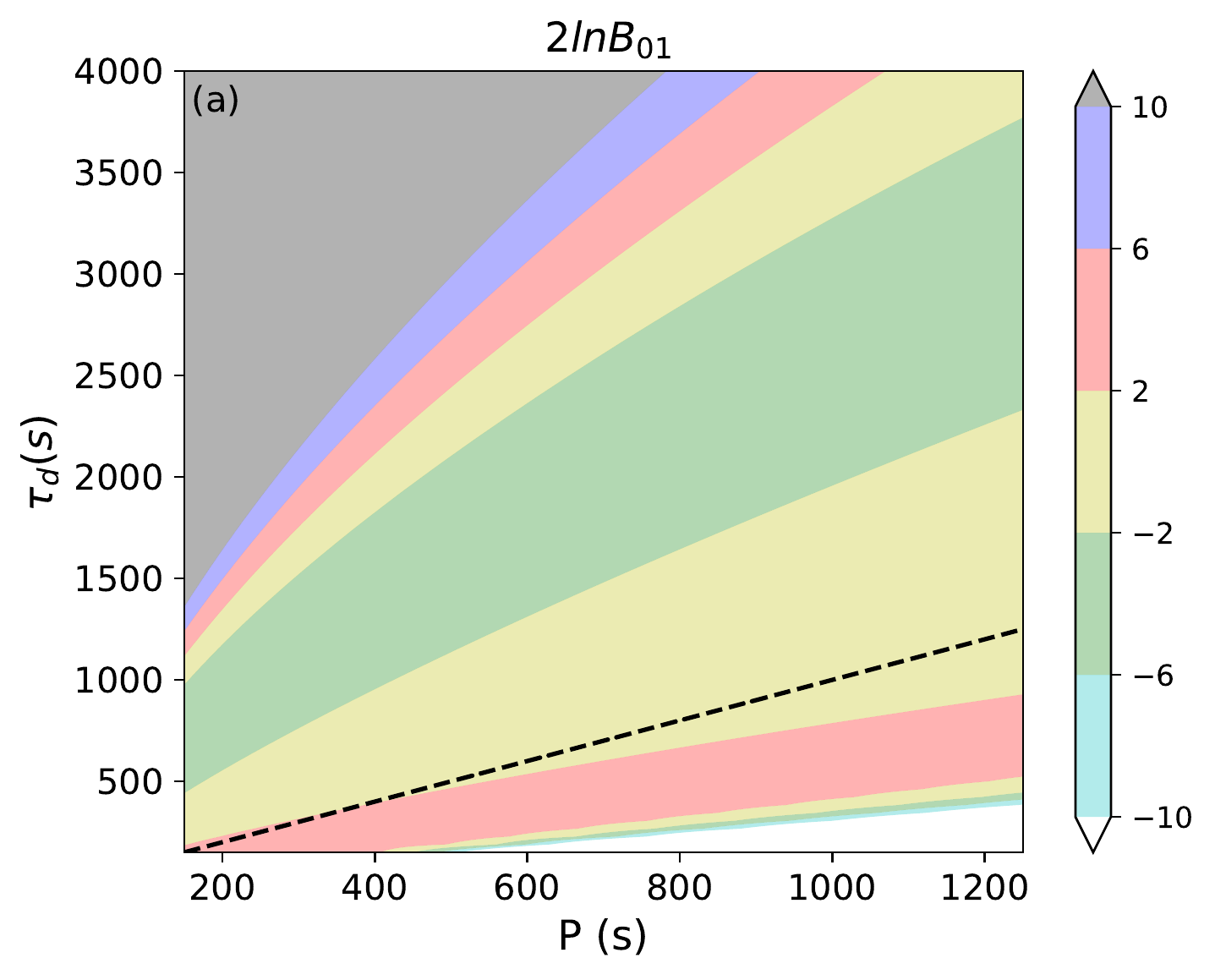}
	\includegraphics[trim={0.4cm 0cm 0cm 1cm},scale=0.44]{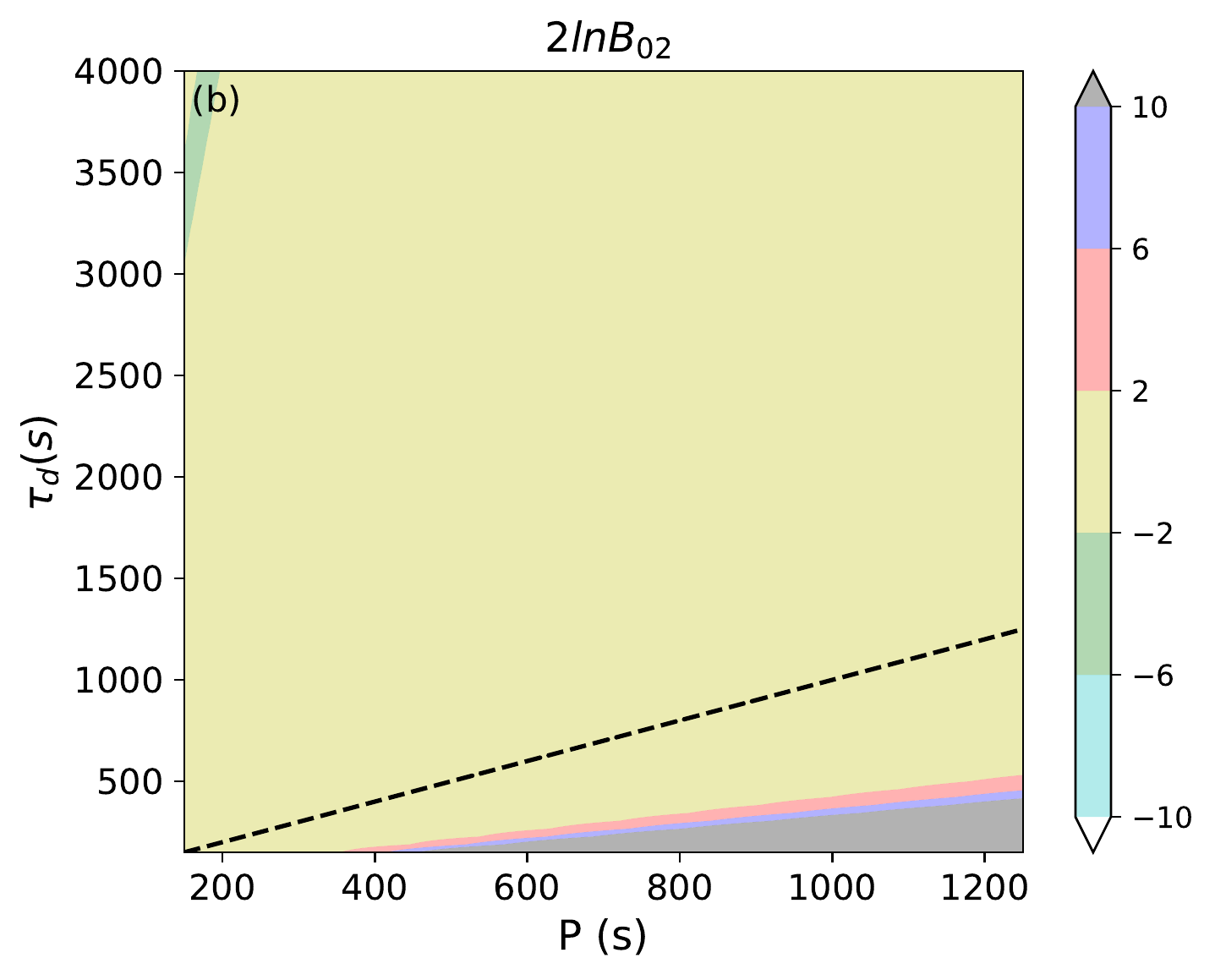}
	\includegraphics[trim={0.4cm 0cm 0.2cm 1cm},scale=0.44]{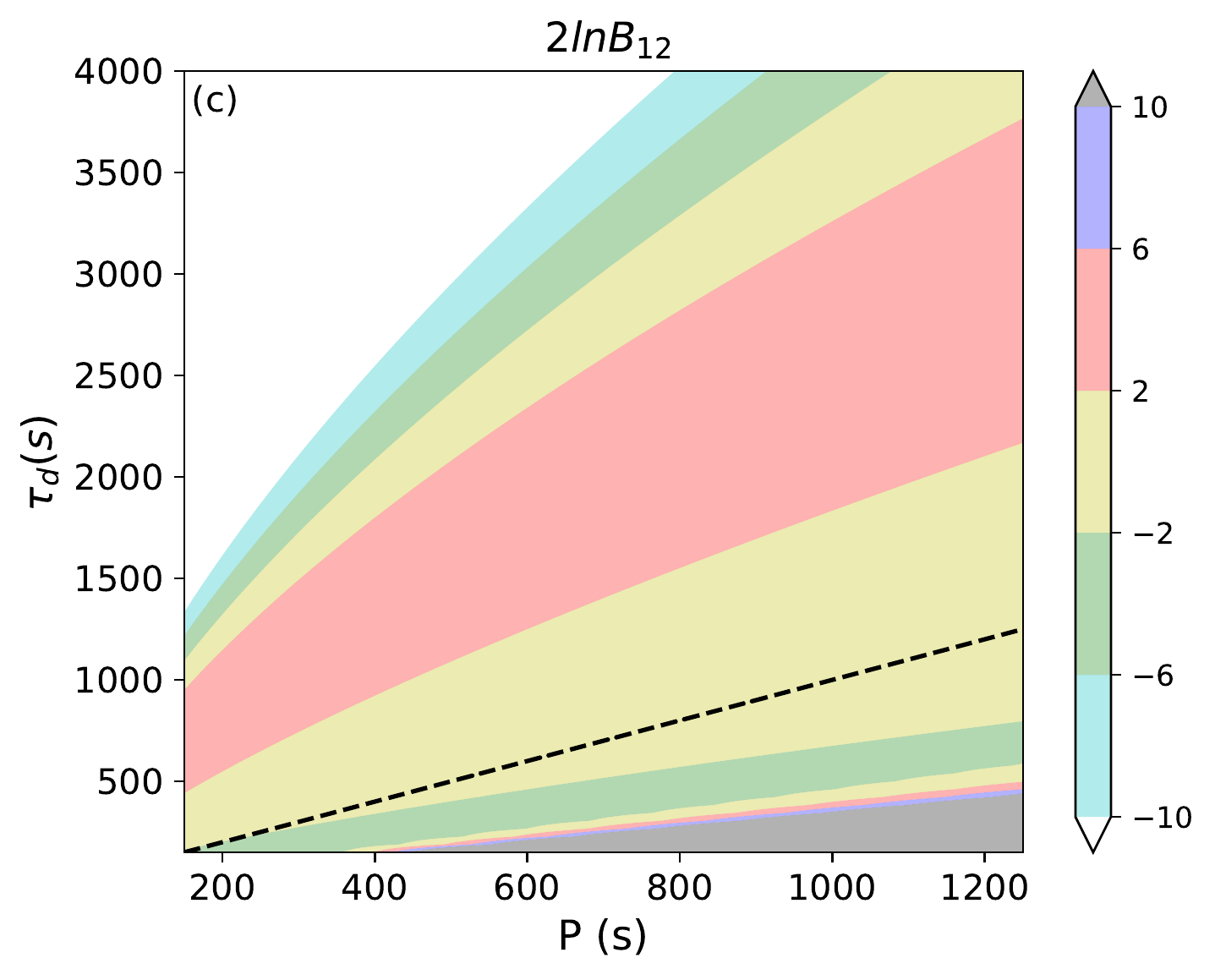}
	\caption{Bayes factors in the one-to-one comparison between resonant absorption, phase mixing, and wave leakage mechanisms as a function of the observables period and damping time with uncertainties of 10\% for each. The dashed lines indicate $\tau_{d}=P$. The different levels of evidence of Table \ref{tab1} are indicated in the color bars. NW (yellow), PE (green/red), SE (blue/purple), VSE (white/gray). \label{fig:f6}}
\end{figure*}

	It is customary to address the comparison between damping mechanism using data on periods and damping times in a scatter plot where these data are then fitted to a straight line, according to the scaling law hypothesis \citep{Ofman2002}. We proceed differently and next evaluate the plausibility between damping mechanisms by calculating the Bayes factor distributions in such a two-dimensional plane with the two observables of interest, period and damping time.  Hence $d=\{P,\tau_{\rm d}\}$. To do so we consider the damping ratios resulting for different combinations of $\tau_{d}$ and $P$ on a wide range of values and a fixed uncertainty. Figure \ref{fig:f6} shows the obtained Bayes factor distributions. The general conclusion is that the evidence for any of the considered models against an alternative depends on the particular combination of observed periods and damping times. 
	
	In particular, for resonant absorption against phase mixing, Figure \ref{fig:f6}a shows that at the upper-left corner of this panel, corresponding to large values of the damping ratio, we obtain strong (purple) and very strong (gray) evidence in favor of resonant absorption. The lower-right corner corresponding to low damping ratios shows very strong evidence (white) in favor of the phase mixing model. In the remaining regions different colored bands denote positive (pink for resonant and green for phase mixing) or insignificant evidence (yellow), depending on the particular combination of the two observables.

	In the comparison between resonant absorption and wave leakage, Figure \ref{fig:f6}b, most of the ($\tau_{\rm d},P$)- plane is colored in yellow resulting in a lack of evidence for a particular model. Grey and purple regions (very strong and strong evidence) indicate the dominance of the resonant mechanism for the lowest values of the damping ratio and positive evidence (green for wave leakage and pink for resonant) is found in very narrow regions of the plane of observables. 
	
	Finally, Figure \ref{fig:f6}c shows the results from the comparison between phase mixing and wave leakage. The white (very strong evidence) and the blue (strong evidence) regions are located at period and damping time combinations corresponding to large damping ratios and point out to the dominance of the wave leakage model over phase mixing. For low damping ratios, we find very strong evidence (gray) in favor of phase mixing. The remaining combinations lead to positive (pink for phase mixing and green for wave leakage) or inconclusive evidence (yellow) depending on the observables.

	In the three panels of Figure \ref{fig:f6}, focusing in observations with values of the damping ratio larger than 1, which are located in the upper-left side of the dashed line, resonant absorption and wave leakage are more plausible than phase mixing but no distinction can be made to favor any of them. Phase mixing seems to be the most plausible mechanism in the intermediate region of the observable ($\tau_{\rm d}, P$)-plane (green color in Figure \ref{fig:f6}a and pink color in Figure \ref{fig:f6}c).
	
	Until now, model comparison is pursued by considering hypothetical values for observed periods, damping times, and their measurement errors. The full potential of the method here presented is discernible when applied to real events of damped transverse loop oscillations. For this reason, we consider a selection of 89 loop oscillation events for which periods and damping times are listed in the databases presented by \citet{Verwichte2013} and \citet{Goddard2016}, discarding events for which errors were not reported. The considered events and their oscillation properties are presented in Table \ref{tabA} of Appendix \ref{apen1}. They are sorted in increasing value for their corresponding damping ratio in order to locate the events easily according to results. We further include for each event the inferred parameters, computed following the methods described in Section \ref{inference results}. 

\begin{figure*}
	\figurenum{7}
	\vspace{0.2cm}
	\centering
	\includegraphics[trim={1.25cm 0cm 0cm 1cm},scale=0.4]{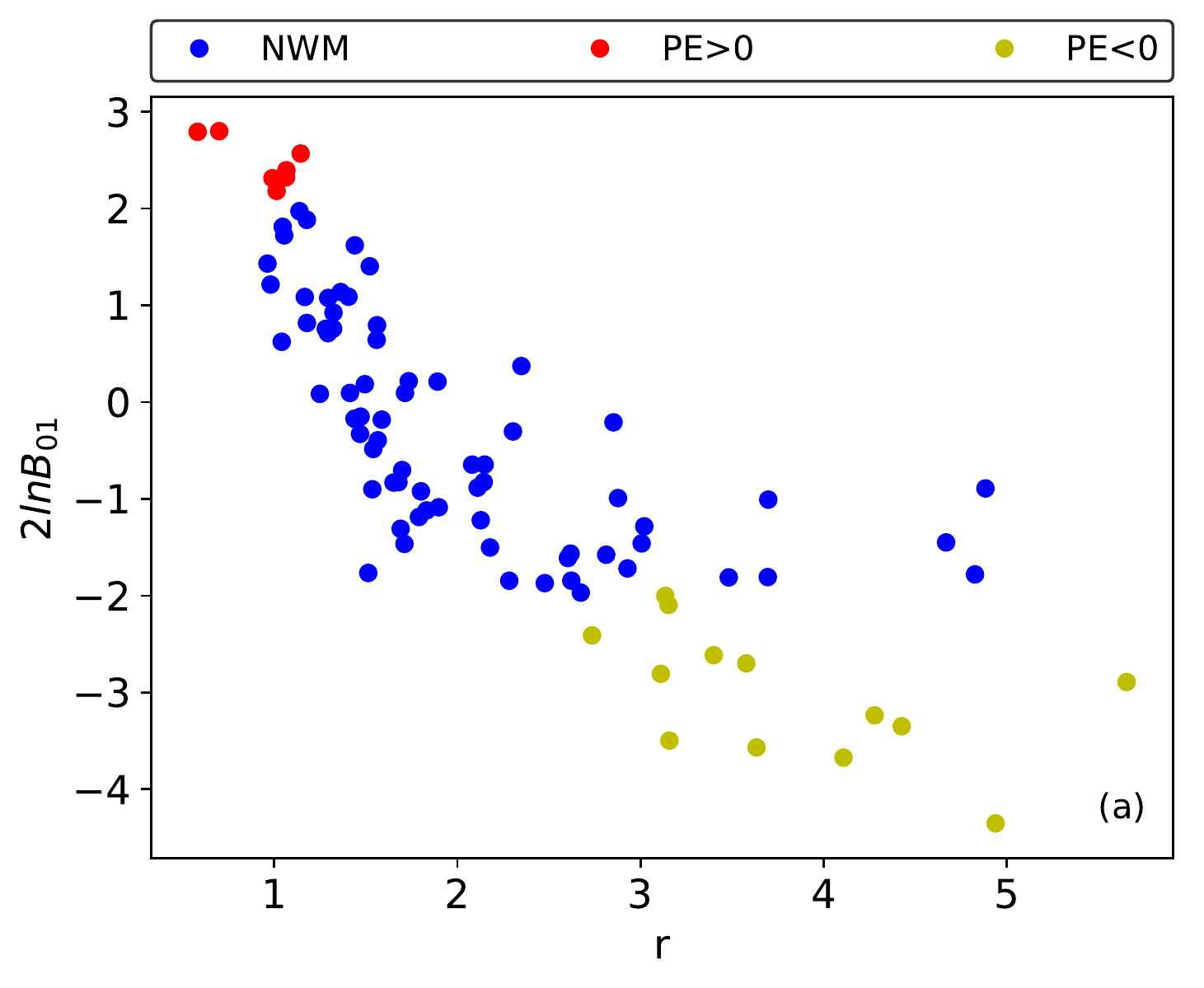}
	\includegraphics[trim={0.4cm 0cm 0cm 1cm},scale=0.4]{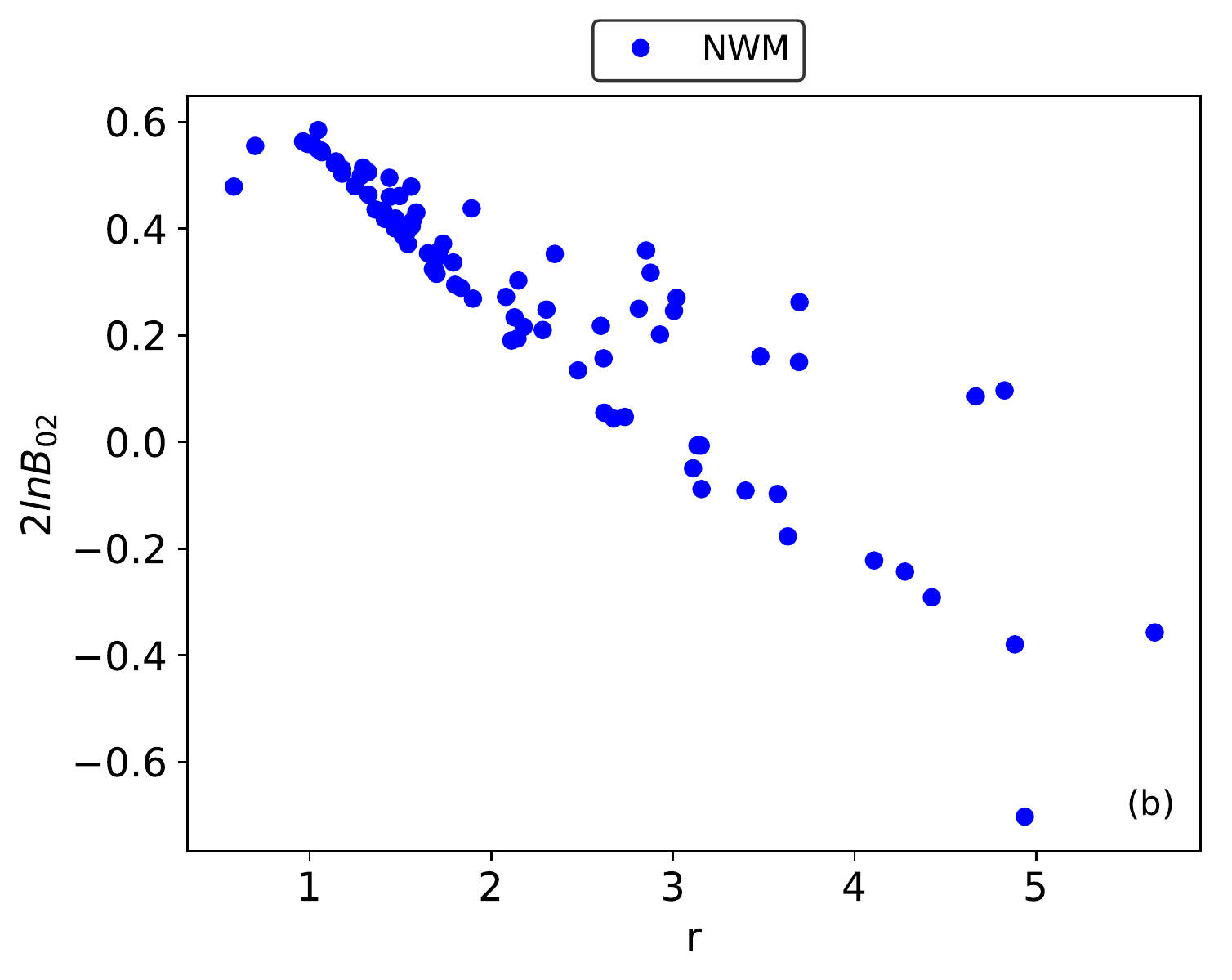}
	\includegraphics[trim={0.4cm 0cm 0.2cm 1cm},scale=0.4]{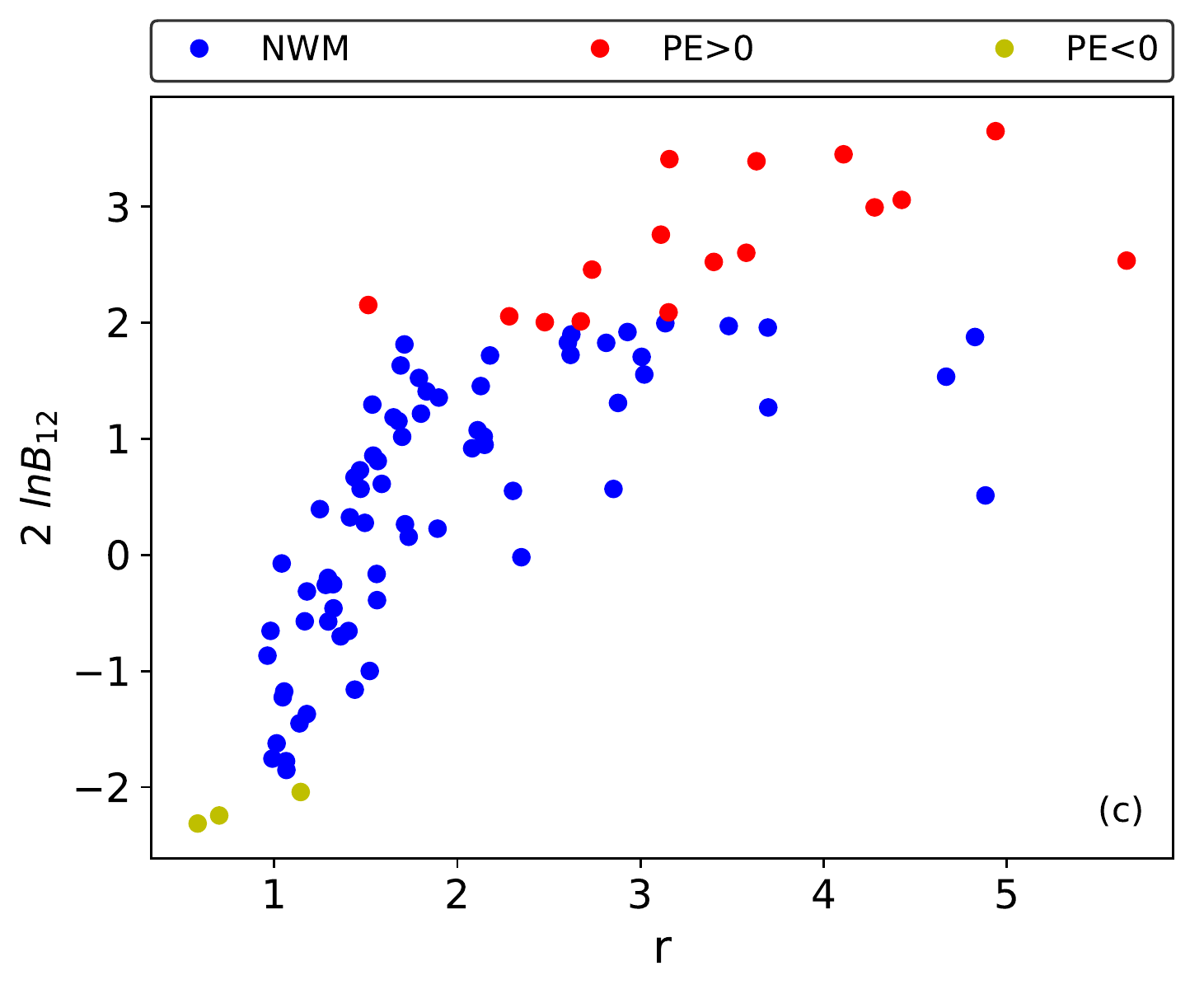}
	\caption{Representation of the Bayes factors computed for the 89 events selected from \citet{Verwichte2013} and \citet{Goddard2016}. The different panels correspond to the three one-to-one comparisons between resonant absorption, phase mixing, and wave leakage, here represented with the subscripts 0, 1, and 2 respectively.\label{fig:f7}}
\end{figure*}

	Figure \ref{fig:f7} shows scatter plots that display Bayes factor and damping ratio values for the selected 89 events for each one-to-one model comparison between our three damping mechanisms. The colors indicate the level of evidence, based on the magnitude of the corresponding Bayes factor. The Bayes factor values are also listed in Table \ref{tabA}. The conspicuous result is that the blue color, which corresponds to absence of evidence for any damping mechanism, dominates in all three panels.
	
	For the comparison between resonant absorption and phase mixing (left panel), in approximately 78\% of the events we have not sufficient evidence to favor one model or the other (NWM). Some events (colored in red and yellow) show positive evidence for either resonant absorption or for phase mixing. The evidence is positive for resonant absorption in 8\% of the events, with strong damping (low values of $r$)
	and positive evidence for phase mixing in about 14\%  of the events, with damping ratio values in between 3 and 6. In the middle panel the analysis for resonant absorption vs. wave leakage is shown. In all events the Bayes factor and therefore the evidence is not strong enough to support any of the two mechanisms. Finally, the right panel shows the comparison between phase mixing and wave leakage. For 79\% of the events the evidence is inconclusive, with the events distributed among the full range of damping ratios. Only for 3\% of the events we obtain positive evidence in favor of wave leakage, for loop oscillations with very strong damping, and for another 18\% positive evidence in favor of phase mixing.	
	
	\section{SUMMARY AND CONCLUSIONS}\label{conclusions}
	
	In the last years, high resolution observations have enabled us to better characterize the damping of transverse oscillations in coronal loops. More accurate estimates of periods and damping times are now obtained and the size of the databases has increased. However, the discussion about the mechanisms involved in the quick damping of the observed oscillations remains latent. The possible mechanisms were proposed a few decades ago but no self-consistent method exists yet to assess which one better accounts for the observed damping phenomenon. The widespread approach has been to resort to the use of scaling laws that would apparently be an intrinsic property of each considered theoretical mechanism. This has proven to be of little use since, physical properties of coronal loops being unknown and probably different, mechanisms such as resonant absorption are known to predict not only one but many scalings with different power indexes.
	
	In this paper, we have presented a method to compute relative probabilities between alternative damping mechanisms for transverse coronal loop oscillations. We considered three mechanisms among those that have been proposed: resonant absorption, phase mixing of Alfv\'en waves, and wave leakage. They all are in principle likely to occur because of the highly inhomogeneous nature of the corona across the oscillating waveguides.  Their direct applicability to the damping of transverse loop oscillations is different, with phase-mixing presenting serious limitations. Together with model comparison, the coronal loop physical parameters that characterize each mechanism have been inferred. The analysis was carried out using Bayesian analysis tools. Bayesian inference enabled us to obtain probability density functions for the parameters of interest, with correct propagation of uncertainty from observables to physical parameters. The computation of marginal likelihoods informs us on the likelihood of each mechanism to generate the observed data. Finally, Bayes factors quantify the level of evidence for one mechanism against another alternative. 
	
	Our inference results indicate that physical parameters such as the density contrast, the transverse density inhomogeneity length-scales, and the aspect ratio of coronal loops can be properly inferred. Considering typically observed damping ratio values, the obtained distributions peak at reasonable values of the unknown parameters. Our model comparison results indicate that, as a general rule, a single damping mechanism cannot explain the observed damping of coronal loop oscillations. However, the method enables us to assign a level of evidence to each considered damping model. Considering hypothetical observed damping ratios over a range of plausible values, we found that resonant absorption and wave leakage offer the most probable explanation in strong damping regimes, while phase mixing is the best candidate for moderate/weak damping. The method was then applied to a large selection of loop oscillation events compiled in databases that provide accurate measurements of period and damping time uncertainties. A frequentist analysis of the obtained Bayes factors indicates that only in very few cases the evidence is large enough to support a particular damping mechanism. For the rest of the cases, nothing can be stated. Since the uncertainty on the measured times-scales is essential when translated into levels of evidence, the future increase in the precision of data becomes relevant in order to determine the associated damping processes.
	
	The results here presented have not given a definitive answer to the question of what mechanism is responsible for the quick damping of coronal loop oscillations, but the method makes use of all the available information - models, observed data with their uncertainty, and prior information - in a consistent manner.  The method is applicable to additional damping models and formulas. Approximate answers have been obtained by considering the pertinent question of quantifying the evidence for each model in view of data, rather than obtaining exact answers to the more misleading question of how well the data fit to questionable theoretical power laws.  

\acknowledgments
We acknowledge financial support from the Spanish Ministry of Economy and Competitiveness (MINECO) through projects AYA2014-55456-P (Bayesian Analysis of the Solar Corona), AYA2014-60476-P (Solar Magnetometry in the Era of Large Telescopes), and from FEDER funds. M.M-S. acknowledges financial support through a Severo Ochoa FPI Fellowship under the project SEV-2011-0187-03. I.A. acknowledges financial support through a Ram\'on y Cajal Fellowship. 


\software{emcee \citep{emcee}
          }



\appendix

\section{ANALYSED CORONAL LOOP DATA}{\label{apen1}}
\begin{longrotatetable}

	\begin{deluxetable}{cccCCCCCCCCCC}
		\tablecaption{Transverse loop oscillations data, inference results and Bayes factors. \label{tabA}}
		\tabletypesize{\footnotesize}
		\tablehead{\colhead{Ref} &\colhead{Ev.ID} & \colhead{Lp.ID} & \colhead{P} & \dcolhead{\tau_d} & \colhead{r} & \colhead{l/R} & \dcolhead{\zeta} & \colhead{w} & \dcolhead{R/L} & \dcolhead{2lnBF_{01}} & \dcolhead{2lnBF_{02}} & \dcolhead{2lnBF_{12}}\\
			\colhead{ } &\colhead{ } &\colhead{ } & \colhead{(s)} & \colhead{(s)} & \colhead{ } & \colhead{ } & \colhead{ } & \colhead{(Mm)} & \colhead{ } & \colhead{ } & \colhead{ } & \colhead{ }}
		\tablecolumns{11}
		\startdata
		1 & 24 & \nodata & 895 \pm 2 & 521 \pm 8 & 0.6 \pm 0.0 & 1.5^{+0.3}_{-0.1} & 6.5^{+2.4}_{-2.2}  & 1.6^{+0.1}_{-0.0} & 0.27^{+0.00}_{-0.00} & 3.45 ^* & 0.48 &-2.97 ^* \\
		2 & 10 & 1 & 687.6 \pm 10.2 & 481.2 \pm 65.4 & 0.7 \pm 0.1 & 1.4^{+0.3}_{-0.2} & 6.1^{+2.6}_{-2.3}  & 1.9^{+0.4}_{-0.3} & 0.25^{+0.02}_{-0.02} & 2.80 ^* & 0.56 & -2.24 ^* \\
		2 & 56 & 7 & 849.6 \pm 33.0 & 818.4 \pm 235.8 & 1.0 \pm 0.3 & 1.2^{+0.5}_{-0.3} & 5.8^{+2.8}_{-2.6}  & 3.4^{+1.5}_{-1.3} & 0.22^{+0.04}_{-0.03} & 1.43 & 0.56 & -0.87 \\
		2 & 48 & 3 & 964.8 \pm 12.6 & 945.6 \pm 185.4 & 1.0 \pm 0.2 & 1.1^{+0.4}_{-0.3} & 5.5^{+3.0}_{-2.5}  & 3.6^{+1.1}_{-1.0} & 0.21^{+0.03}_{-0.02} & 1.21 & 0.56 & -0.65 \\
		2 & 9 & 1 & 308.4 \pm 10.2 & 305.4 \pm 58.8 & 1.0 \pm 0.2 & 1.1^{+0.4}_{-0.3} & 5.5^{+3.0}_{-2.5}  & 2.2^{+0.6}_{-0.6} & 0.21^{+0.03}_{-0.02} & 2.31 ^* & 0.56 & -1.75 \\
		1 & 40 & \nodata & 302 \pm 14 & 306 \pm 43 & 1.0 \pm 0.1 & 1.0^{+0.4}_{-0.2} & 5.3^{+3.1}_{-2.5}  & 2.2^{+0.5}_{-0.4} & 0.21^{+0.02}_{-0.02} & 2.18 ^* & 0.56 & -1.62 \\
		1 & 44 & \nodata & 1170 \pm 6 & 1218 \pm 48 & 1.0 \pm 0.0 & 0.9^{+0.5}_{-0.1} & 5.2^{+3.2}_{-2.5}  & 4.2^{+0.3}_{-0.2} & 0.20^{+0.00}_{-0.00} & 0.62 & 0.55 & -0.07 \\
		1 & 25 & \nodata & 452 \pm 1 & 473 \pm 6 & 1.0 \pm 0.0 & 0.9^{+0.5}_{-0.1} & 5.2^{+3.2}_{-2.5}  & 2.8^{+0.0}_{-0.1} & 0.20^{+0.00}_{-0.00} & 1.57 & 0.55 & -1.02 \\
		2 & 43 & 1 & 428.4 \pm 4.2 & 451.8 \pm 87.0 & 1.1 \pm 0.2 & 1.0^{+0.5}_{-0.2} & 5.4^{+3.1}_{-2.5}  & 2.8^{+0.8}_{-0.7} & 0.21^{+0.03}_{-0.02} & 1.72 & 0.55 & -1.18 \\
		1 & 32 & \nodata & 216 \pm 27 & 230 \pm 23 & 1.1 \pm 0.2 & 1.0^{+0.4}_{-0.2} & 5.3^{+3.2}_{-2.5}  & 2.0^{+0.5}_{-0.4} & 0.20^{+0.02}_{-0.02} & 2.32 ^* & 0.55 & -1.78 \\
		1 & 29 & \nodata & 225 \pm 40 & 240 \pm 45 & 1.1 \pm 0.3 & 1.1^{+0.5}_{-0.3} & 5.6^{+3.0}_{-2.6}  & 2.1^{+0.8}_{-0.7} & 0.21^{+0.04}_{-0.02} & 2.40 ^* & 0.54 & -1.85 \\
		2 & 3 & 2 & 217.2 \pm 4.8 & 247.2 \pm 28.2 & 1.1 \pm 0.1 & 0.9^{+0.4}_{-0.2} & 5.1^{+3.3}_{-2.5}  & 2.2^{+0.4}_{-0.3} & 0.19^{+0.01}_{-0.01} & 1.97 & 0.52 & -1.45 \\
		2 & 16 & 2 & 141.0 \pm 4.2 & 161.4 \pm 38.4 & 1.1 \pm 0.3 & 1.0^{+0.5}_{-0.3} & 5.4^{+3.1}_{-2.6}  & 1.9^{+0.6}_{-0.6} & 0.20^{+0.03}_{-0.02} & 2.57 ^* & 0.53 & -2.04 ^* \\
		2 & 49 & 5 & 481.8 \pm 10.8 & 562.2 \pm 73.2 & 1.2 \pm 0.2 & 0.9^{+0.5}_{-0.2} & 5.1^{+3.3}_{-2.5}  & 3.3^{+0.6}_{-0.6} & 0.19^{+0.02}_{-0.01} & 1.09 & 0.51 & -0.57 \\
		1 & 31 & \nodata & 213 \pm 9 & 251 \pm 36 & 1.2 \pm 0.2 & 0.9^{+0.5}_{-0.2} & 5.1^{+3.3}_{-2.5}  & 2.3^{+0.5}_{-0.5} & 0.19^{+0.02}_{-0.01} & 1.88 & 0.51 & -1.37 \\
		1 & 41 & \nodata & 565 \pm 4 & 666 \pm 42 & 1.2 \pm 0.1 & 0.8^{+0.5}_{-0.1} & 5.1^{+3.3}_{-2.5}  & 3.6^{+0.3}_{-0.4} & 0.19^{+0.01}_{-0.01} & 0.82 & 0.50 & -0.31 \\
		2 & 23 & 1 & 921.6 \pm 24.0 & 1151.4 \pm 93.0 & 1.2 \pm 0.1 & 0.8^{+0.5}_{-0.1} & 5.0^{+3.3}_{-2.5}  & 5.0^{+0.6}_{-0.6} & 0.18^{+0.01}_{-0.01} & 0.09 & 0.48 & 0.39 \\
		2 & 18 & 2 & 571.2 \pm 6.6 & 732.0 \pm 208.2 & 1.3 \pm 0.4 & 1.0^{+0.5}_{-0.3} & 5.4^{+3.1}_{-2.6}  & 4.3^{+1.9}_{-1.6} & 0.19^{+0.04}_{-0.03} & 0.76 & 0.50 & -0.26 \\
		1 & 34 & \nodata & 596 \pm 50 & 771 \pm 336 & 1.3 \pm 0.6 & 1.1^{+0.6}_{-0.4} & 5.7^{+2.9}_{-2.8}  & 4.9^{+2.9}_{-2.4} & 0.20^{+0.06}_{-0.04} & 0.71 & 0.51 & -0.20 \\
		2 & 44 & 3 & 417.0 \pm 8.4 & 540.0 \pm 180.0 & 1.3 \pm 0.4 & 1.0^{+0.6}_{-0.4} & 5.5^{+3.1}_{-2.7}  & 3.8^{+1.9}_{-1.6} & 0.20^{+0.05}_{-0.03} & 1.08 & 0.50 & -0.57 \\
		2 & 9 & 2 & 537.0 \pm 8.4 & 709.8 \pm 285.6 & 1.3 \pm 0.5 & 1.1^{+0.6}_{-0.4} & 5.6^{+3.0}_{-2.8}  & 4.7^{+2.6}_{-2.2} & 0.20^{+0.05}_{-0.04} & 0.76 & 0.51 & -0.25 \\
		2 & 40 & 3 & 331.8 \pm 2.4 & 439.2 \pm 64.8 & 1.3 \pm 0.2 & 0.8^{+0.5}_{-0.2} & 5.0^{+3.3}_{-2.5}  & 3.3^{+0.7}_{-0.7} & 0.18^{+0.02}_{-0.01} & 0.92 & 0.46 & -0.46 \\
		1 & 30 & \nodata & 215 \pm 5 & 293 \pm 18 & 1.4 \pm 0.1 & 0.7^{+0.5}_{-0.1} & 4.9^{+3.4}_{-2.5}  & 2.8^{+0.3}_{-0.2} & 0.18^{+0.01}_{-0.00} & 1.14 & 0.44 & -0.70 \\
		1 & 35 & \nodata & 212 \pm 20 & 298 \pm 30 & 1.4 \pm 0.2 & 0.8^{+0.5}_{-0.2} & 4.9^{+3.4}_{-2.5}  & 2.9^{+0.6}_{-0.6} & 0.18^{+0.01}_{-0.01} & 1.09 & 0.43 & -0.66 \\
		1 & 33 & \nodata & 520 \pm 5 & 735 \pm 53 & 1.4 \pm 0.1 & 0.7^{+0.5}_{-0.1} & 4.8^{+3.4}_{-2.5}  & 4.5^{+0.5}_{-0.5} & 0.17^{+0.01}_{-0.01} & 0.09 & 0.42 & 0.32 \\
		2 & 48 & 1 & 916.8 \pm 9.6 & 1318.8 \pm 936.0 & 1.4 \pm 1.0 & 1.1^{+0.6}_{-0.5} & 5.8^{+2.9}_{-2.8}  & 8.2^{+6.0}_{-4.8} & 0.20^{+0.06}_{-0.05} & -0.17 & 0.50 & 0.67 \\
		1 & 46 & \nodata & 150 \pm 5 & 216 \pm 60 & 1.4 \pm 0.4 & 0.9^{+0.6}_{-0.3} & 5.2^{+3.2}_{-2.6}  & 2.7^{+1.1}_{-0.9} & 0.18^{+0.04}_{-0.02} & 1.62 & 0.46 & -1.16 \\
		2 & 19 & 2 & 676.2 \pm 7.2 & 993.0 \pm 86.4 & 1.5 \pm 0.1 & 0.7^{+0.5}_{-0.1} & 4.8^{+3.5}_{-2.5}  & 5.4^{+0.8}_{-0.7} & 0.17^{+0.01}_{-0.01} & -0.33 & 0.40 & 0.73 \\
		2 & 49 & 4 & 627.0 \pm 10.2 & 922.8 \pm 154.8 & 1.5 \pm 0.2 & 0.8^{+0.5}_{-0.2} & 4.9^{+3.4}_{-2.5}  & 5.3^{+1.4}_{-1.2} & 0.17^{+0.02}_{-0.01} & -0.15 & 0.42 & 0.57 \\
		2 & 44 & 2 & 586.8 \pm 11.4 & 877.2 \pm 297.6 & 1.5 \pm 0.5 & 0.9^{+0.6}_{-0.4} & 5.3^{+3.2}_{-2.7}  & 5.6^{+2.8}_{-2.3} & 0.19^{+0.05}_{-0.03} & 0.19 & 0.46 & 0.28 \\
		1 & 27 & \nodata & 2418 \pm 5 & 3660 \pm 80 & 1.5 \pm 0.0 & 0.7^{+0.5}_{-0.1} & 4.8^{+3.5}_{-2.5}  & 10.6^{+0.4}_{-0.3} & 0.17^{+0.00}_{-0.00} & -1.78 & 0.38 & 2.15 ^* \\
		1 & 38 & \nodata & 115 \pm 2 & 1750 \pm 30 & 1.5 \pm 0.3 & 0.7^{+0.5}_{-0.2} & 4.8^{+3.4}_{-2.5}  & 2.5^{+0.6}_{-0.6} & 0.17^{+0.02}_{-0.02} & 1.40 & 0.40 & -1.00 \\
		2 & 24 & 1 & 1071.6 \pm 18.0 & 1645.8 \pm 255.6 & 1.5 \pm 0.2 & 0.7^{+0.5}_{-0.2} & 4.8^{+3.5}_{-2.5}  & 7.4^{+1.8}_{-1.6} & 0.17^{+0.02}_{-0.01} & -0.90 & 0.39 & 1.29 \\
		1 & 45 & \nodata & 623 \pm 4 & 960 \pm 60 & 1.5 \pm 0.1 & 0.7^{+0.5}_{-0.1} & 4.8^{+3.5}_{-2.5}  & 5.6^{+0.5}_{-0.5} & 0.16^{+0.01}_{-0.00} & -0.48 & 0.37 & 0.86 \\
		2 & 25 & 1 & 307.8 \pm 6.6 & 480.0 \pm 300.0 & 1.6 \pm 1.0 & 1.1^{+0.6}_{-0.5} & 5.7^{+2.9}_{-2.8}  & 5.4^{+4.1}_{-2.9} & 0.20^{+0.06}_{-0.05} & 0.64 & 0.48 & -0.16 \\
		2 & 1 & 1 & 205.2 \pm 3.6 & 320.4 \pm 67.2 & 1.6 \pm 0.3 & 0.8^{+0.5}_{-0.2} & 4.9^{+3.4}_{-2.5}  & 3.4^{+1.1}_{-1.0} & 0.17^{+0.02}_{-0.02} & 0.79 & 0.40 & -0.39 \\
		2 & 26 & 1 & 717.0 \pm 7.8 & 1122.6 \pm 270.0 & 1.6 \pm 0.4 & 0.8^{+0.5}_{-0.2} & 5.0^{+3.3}_{-2.6}  & 6.4^{+2.3}_{-2.1} & 0.17^{+0.03}_{-0.02} & -0.39 & 0.41 & 0.81 \\
		1 & 26 & \nodata & 630 \pm 30 & 1000 \pm 300 & 1.6 \pm 0.5 & 0.9^{+0.6}_{-0.3} & 5.2^{+3.3}_{-2.6}  & 6.3^{+2.8}_{-2.4} & 0.18^{+0.05}_{-0.03} & -0.18 & 0.43 & 0.61 \\
		2 & 56 & 2 & 712.8 \pm 7.8 & 1177.2 \pm 177.6 & 1.7 \pm 0.2 & 0.7^{+0.5}_{-0.2} & 4.8^{+3.5}_{-2.5}  & 6.7^{+1.6}_{-1.4} & 0.16^{+0.02}_{-0.01} & -0.83 & 0.35 & 1.18 \\
		2 & 34 & 1 & 597.0 \pm 16.2 & 1002.0 \pm 61.8 & 1.7 \pm 0.1 & 0.6^{+0.5}_{-0.1} & 4.7^{+3.5}_{-2.5}  & 6.2^{+0.7}_{-0.6} & 0.16^{+0.01}_{-0.00} & -0.83 & 0.32 & 1.15 \\
		2 & 48 & 2 & 945.6 \pm 7.2 & 1598.4 \pm 130.2 & 1.7 \pm 0.1 & 0.6^{+0.5}_{-0.1} & 4.7^{+3.5}_{-2.5}  & 7.8^{+1.0}_{-0.9} & 0.16^{+0.01}_{-0.01} & -1.31 & 0.32 & 1.63 \\
		1 & 50 & \nodata & 491 \pm 18 & 834 \pm 6 & 1.7 \pm 0.1 & 0.6^{+0.5}_{-0.1} & 4.6^{+3.6}_{-2.5}  & 5.7^{+0.4}_{-0.3} & 0.16^{+0.00}_{-0.00} & -0.70 & 0.31 & 1.02 \\
		2 & 24 & 3 & 1227.6 \pm 34.8 & 2100.6 \pm 386.4 & 1.7 \pm 0.3 & 0.7^{+0.5}_{-0.2} & 4.8^{+3.5}_{-2.5}  & 9.3^{+2.6}_{-2.4} & 0.16^{+0.02}_{-0.02} & -1.46 & 0.35 & 1.81 \\
		1 & 48 & \nodata & 273 \pm 54 & 468 \pm 36 & 1.7 \pm 0.4 & 0.7^{+0.5}_{-0.2} & 4.8^{+3.5}_{-2.5}  & 4.5^{+1.4}_{-1.3} & 0.16^{+0.02}_{-0.02} & 0.09 & 0.36 & 0.26 \\
		1 & 36 & \nodata & 256 \pm 22 & 444 \pm 105 & 1.7 \pm 0.4 & 0.7^{+0.6}_{-0.2} & 4.9^{+3.4}_{-2.6}  & 4.5^{+1.7}_{-1.5} & 0.16^{+0.03}_{-0.02} & 0.22 & 0.37 & 0.16 \\
		2 & 56 & 5 & 810.0 \pm 9.6 & 1450.2 \pm 307.8 & 1.8 \pm 0.4 & 0.7^{+0.5}_{-0.2} & 4.8^{+3.5}_{-2.5}  & 8.1^{+2.6}_{-2.3} & 0.16^{+0.02}_{-0.02} & -1.19 & 0.34 & 1.52 \\
		2 & 43 & 3 & 501.0 \pm 4.8 & 902.4 \pm 108.6 & 1.8 \pm 0.2 & 0.6^{+0.5}_{-0.1} & 4.6^{+3.6}_{-2.5}  & 6.4^{+1.2}_{-1.1} & 0.15^{+0.01}_{-0.01} & -0.92 & 0.29 & 1.22 \\
		2 & 31 & 2 & 575.4 \pm 5.4 & 1054.2 \pm 141.0 & 1.8 \pm 0.2 & 0.6^{+0.5}_{-0.1} & 4.6^{+3.6}_{-2.5}  & 7.0^{+1.5}_{-1.3} & 0.15^{+0.01}_{-0.01} & -1.12 & 0.29 & 1.41 \\
		1 & 42 & \nodata & 222 \pm 18 & 420 \pm 360 & 1.9 \pm 1.6 & 1.1^{+0.6}_{-0.6} & 5.7^{+2.9}_{-2.9}  & 7.2^{+5.9}_{-4.3} & 0.19^{+0.07}_{-0.06} & 0.21 & 0.44 & 0.23 \\
		1 & 43 & \nodata & 474 \pm 12 & 900 \pm 120 & 1.9 \pm 0.3 & 0.6^{+0.5}_{-0.1} & 4.6^{+3.6}_{-2.5}  & 6.8^{+1.4}_{-1.4} & 0.15^{+0.01}_{-0.01} & -1.09 & 0.27 & 1.36 \\
		2 & 40 & 8 & 259.8 \pm 4.8 & 540.6 \pm 129.6 & 2.1 \pm 0.5 & 0.6^{+0.5}_{-0.2} & 4.7^{+3.6}_{-2.5}  & 5.9^{+2.2}_{-1.9} & 0.15^{+0.03}_{-0.02} & -0.65 & 0.27 & 0.92 \\
		2 & 29 & 1 & 222.6 \pm 3.0 & 469.8 \pm 37.2 & 2.1 \pm 0.2 & 0.5^{+0.5}_{-0.1} & 4.4^{+3.7}_{-2.5}  & 5.4^{+0.6}_{-0.7} & 0.14^{+0.01}_{-0.01} & -0.89 & 0.19 & 1.07 \\
		2 & 40 & 9 & 370.8 \pm 3.0 & 789.0 \pm 159.6 & 2.1 \pm 0.4 & 0.6^{+0.5}_{-0.2} & 4.6^{+3.6}_{-2.5}  & 7.2^{+2.2}_{-2.0} & 0.14^{+0.02}_{-0.01} & -1.22 & 0.23 & 1.45 \\
		2 & 4 & 2 & 208.2 \pm 1.8 & 446.4 \pm 60.0 & 2.1 \pm 0.3 & 0.5^{+0.5}_{-0.1} & 4.5^{+3.7}_{-2.5}  & 5.4^{+1.1}_{-1.1} & 0.14^{+0.01}_{-0.01} & -0.83 & 0.19 & 1.02 \\
		1 & 49 & \nodata & 282 \pm 6 & 606 \pm 186 & 2.1 \pm 0.7 & 0.7^{+0.6}_{-0.3} & 4.9^{+3.4}_{-2.7}  & 6.6^{+3.0}_{-2.5} & 0.15^{+0.04}_{-0.02} & -0.65 & 0.30 & 0.95 \\
		2 & 44 & 1 & 433.8 \pm 3.6 & 945.0 \pm 185.4 & 2.2 \pm 0.4 & 0.6^{+0.5}_{-0.2} & 4.5^{+3.6}_{-2.5}  & 8.0^{+2.4}_{-2.2} & 0.14^{+0.02}_{-0.01} & -1.50 & 0.22 & 1.72 \\
		2 & 56 & 1 & 544.2 \pm 8.4 & 1242.6 \pm 282.6 & 2.3 \pm 0.5 & 0.6^{+0.5}_{-0.2} & 4.6^{+3.6}_{-2.5}  & 9.7^{+3.3}_{-2.9} & 0.14^{+0.02}_{-0.02} & -1.85 & 0.21 & 2.06 ^* \\
		1 & 37 & \nodata & 135 \pm 9 & 311 \pm 85 & 2.3 \pm 0.6 & 0.6^{+0.6}_{-0.2} & 4.7^{+3.5}_{-2.6}  & 5.0^{+2.2}_{-1.8} & 0.15^{+0.04}_{-0.02} & -0.30 & 0.25 & 0.55 \\
		1 & 39 & \nodata & 103 \pm 8 & 242 \pm 114 & 2.3 \pm 1.1 & 0.9^{+0.7}_{-0.4} & 5.3^{+3.2}_{-2.8}  & 5.1^{+3.3}_{-2.5} & 0.17^{+0.07}_{-0.04} & 0.37 & 0.35 & -0.02 \\
		2 & 40 & 7 & 343.2 \pm 3.6 & 850.2 \pm 163.8 & 2.5 \pm 0.5 & 0.5^{+0.5}_{-0.1} & 4.4^{+3.7}_{-2.5}  & 8.6^{+2.4}_{-2.3} & 0.13^{+0.02}_{-0.01} & -1.87 & 0.13 & 2.00 ^* \\
		1 & 51 & \nodata & 348 \pm 7 & 906 \pm 288 & 2.6 \pm 0.8 & 0.6^{+0.7}_{-0.3} & 4.8^{+3.5}_{-2.7}  & 9.7^{+4.4}_{-3.8} & 0.14^{+0.05}_{-0.02} & -1.61 & 0.22 & 1.83 \\
		2 & 1 & 2 & 246.6 \pm 3.0 & 645.6 \pm 167.4 & 2.6 \pm 0.7 & 0.5^{+0.6}_{-0.2} & 4.5^{+3.6}_{-2.5}  & 8.1^{+3.1}_{-2.7} & 0.13^{+0.03}_{-0.02} & -1.57 & 0.16 & 1.72 \\
		2 & 43 & 2 & 216.0 \pm 1.8 & 566.4 \pm 55.2 & 2.6 \pm 0.3 & 0.4^{+0.5}_{-0.1} & 4.3^{+3.8}_{-2.4}  & 7.3^{+1.1}_{-1.0} & 0.13^{+0.01}_{-0.00} & -1.85 & 0.05 & 1.90 \\
		2 & 8 & 1 & 224.4 \pm 4.2 & 600.0 \pm 60.0 & 2.7 \pm 0.3 & 0.4^{+0.5}_{-0.1} & 4.3^{+3.8}_{-2.4}  & 7.7^{+1.2}_{-1.1} & 0.13^{+0.01}_{-0.01} & -1.97 & 0.04 & 2.01 ^* \\
		1 & 52 & \nodata & 340 \pm 3 & 930 \pm 144 & 2.7 \pm 0.4 & 0.4^{+0.5}_{-0.1} & 4.3^{+3.8}_{-2.4}  & 9.8^{+2.4}_{-2.1} & 0.13^{+0.01}_{-0.01} & -2.41 ^* & 0.05 & 2.46 ^* \\
		1 & 13 & \nodata & 448 \pm 18 & 1260 \pm 500 & 2.8 \pm 1.1 & 0.7^{+0.7}_{-0.4} & 5.0^{+3.3}_{-2.8}  & 11.6^{+5.2}_{-5.2} & 0.15^{+0.07}_{-0.03} & -1.58 & 0.25 & 1.83 \\
		1 & 47 & \nodata & 122 \pm 6 & 348 \pm 360 & 2.9 \pm 3.0 & 1.0^{+0.7}_{-0.6} & 5.6^{+3.0}_{-2.9}  & 9.1^{+6.7}_{-5.6} & 0.18^{+0.08}_{-0.07} & -0.21 & 0.36 & 0.57 \\
		1 & 16 & \nodata & 358 \pm 30 & 1030 \pm 570 & 2.9 \pm 1.6 & 0.9^{+0.7}_{-0.5} & 5.3^{+3.2}_{-2.9}  & 10.9^{+5.8}_{-5.8} & 0.16^{+0.08}_{-0.05} & -0.99 & 0.32 & 1.31 \\
		2 & 38 & 2 & 312.0 \pm 4.8 & 913.8 \pm 330.0 & 2.9 \pm 1.1 & 0.7^{+0.7}_{-0.3} & 4.8^{+3.5}_{-2.7}  & 10.8^{+5.0}_{-4.6} & 0.14^{+0.06}_{-0.03} & -1.72 & 0.20 & 1.92 \\
		1 & 17 & \nodata & 326 \pm 45 & 980 \pm 400 & 3.0 \pm 1.3 & 0.7^{+0.8}_{-0.4} & 5.1^{+3.3}_{-2.8}  & 11.2^{+5.3}_{-5.3} & 0.15^{+0.08}_{-0.03} & -1.46 & 0.25 & 1.71 \\
		2 & 20 & 1 & 321.6 \pm 13.8 & 971.4 \pm 460.2 & 3.0 \pm 1.4 & 0.8^{+0.7}_{-0.4} & 5.2^{+3.3}_{-2.8}  & 11.1^{+5.5}_{-5.5} & 0.15^{+0.08}_{-0.04} & -1.28 & 0.27 & 1.55 \\
		2 & 43 & 5 & 270.0 \pm 1.2 & 840.0 \pm 120.0 & 3.1 \pm 0.4 & 0.4^{+0.5}_{-0.1} & 4.2^{+3.8}_{-2.4}  & 10.6^{+2.3}_{-2.2} & 0.12^{+0.01}_{-0.01} & -2.80 ^* & -0.05 & 2.76 ^* \\
		2 & 4 & 1 & 137.4 \pm 1.8 & 430.8 \pm 90.0 & 3.1 \pm 0.7 & 0.4^{+0.5}_{-0.1} & 4.3^{+3.8}_{-2.5}  & 7.8^{+2.4}_{-2.3} & 0.12^{+0.02}_{-0.01} & -2.00 & -0.01 & 2.00 \\
		2 & 45 & 1 & 148.8 \pm 2.4 & 469.2 \pm 99.6 & 3.2 \pm 0.7 & 0.4^{+0.5}_{-0.1} & 4.3^{+3.8}_{-2.5}  & 8.2^{+2.5}_{-2.4} & 0.12^{+0.02}_{-0.01} & -2.10 ^* & -0.01 & 2.09 ^* \\
		2 & 31 & 1 & 460.2 \pm 2.4 & 1453.2 \pm 121.2 & 3.2 \pm 0.3 & 0.4^{+0.5}_{-0.1} & 4.1^{+3.9}_{-2.4}  & 14.0^{+1.8}_{-1.8} & 0.12^{+0.00}_{-0.00} & -3.49 ^* & -0.08 & 3.41 ^* \\
		2 & 11 & 3 & 156.0 \pm 3.0 & 530.4 \pm 90.0 & 3.4 \pm 0.6 & 0.4^{+0.5}_{-0.1} & 4.1^{+3.9}_{-2.4}  & 9.3^{+2.4}_{-2.3} & 0.11^{+0.01}_{-0.01} & -2.61 ^* & -0.09 & 2.52 ^* \\
		1 & 15 & \nodata & 382 \pm 12 & 1330 \pm 528 & 3.5 \pm 1.4 & 0.7^{+0.8}_{-0.3} & 4.9^{+3.4}_{-2.8}  & 12.8^{+4.8}_{-5.7} & 0.13^{+0.07}_{-0.03} & -1.81 & 0.16 & 1.97 \\
		2 & 3 & 1 & 147.6 \pm 1.8 & 528.0 \pm 108.0 & 3.6 \pm 0.7 & 0.4^{+0.5}_{-0.1} & 4.1^{+3.9}_{-2.4}  & 9.8^{+3.0}_{-2.7} & 0.11^{+0.02}_{-0.01} & -2.70 ^* & -0.10 & 2.60 ^* \\
		2 & 32 & 1 & 256.8 \pm 1.2 & 933.0 \pm 73.2 & 3.6 \pm 0.3 & 0.3^{+0.5}_{-0.1} & 4.0^{+3.9}_{-2.4}  & 12.8^{+1.6}_{-1.5} & 0.11^{+0.00}_{-0.00} & -3.57 ^* & -0.18 & 3.39 ^* \\
		1 & 12 & \nodata & 249 \pm 33 & 920 \pm 360 & 3.7 \pm 1.5 & 0.7^{+0.8}_{-0.4} & 4.9^{+3.5}_{-2.8}  & 12.2^{+5.1}_{-5.6} & 0.13^{+0.08}_{-0.03} & -1.81 & 0.15 & 1.96 \\
		1 & 18 & \nodata & 357 \pm 89 & 1320 \pm 720 & 3.7 \pm 2.2 & 0.9^{+0.7}_{-0.5} & 5.3^{+3.2}_{-2.9}  & 11.9^{+5.6}_{-6.4} & 0.16^{+0.09}_{-0.05} & -1.01 & 0.26 & 1.27 \\
		2 & 54 & 5 & 288.0 \pm 6.0 & 1183.2 \pm 193.8 & 4.1 \pm 0.7 & 0.3^{+0.5}_{-0.1} & 4.0^{+3.9}_{-2.4}  & 15.6^{+2.8}_{-3.3} & 0.10^{+0.01}_{-0.01} & -3.68 ^* & -0.23 & 3.45 ^* \\
		2 & 7 & 1 & 101.4 \pm 1.2 & 433.8 \pm 78.0 & 4.3 \pm 0.8 & 0.3^{+0.5}_{-0.1} & 4.0^{+3.9}_{-2.4}  & 10.6^{+2.8}_{-2.6} & 0.10^{+0.01}_{-0.01} & -3.24 ^* & -0.25 & 2.99 ^* \\
		2 & 40 & 2 & 336.6 \pm 1.8 & 1489.8 \pm 204.6 & 4.4 \pm 0.6 & 0.3^{+0.5}_{-0.1} & 3.9^{+4.0}_{-2.3}  & 17.4^{+2.0}_{-2.7} & 0.10^{+0.01}_{-0.01} & -3.35 ^* & -0.30 & 3.06 ^* \\
		1 & 14 & \nodata & 392 \pm 31 & 1830 \pm 790 & 4.7 \pm 2.0 & 0.6^{+0.8}_{-0.4} & 4.9^{+3.5}_{-2.8}  & 13.5^{+4.6}_{-6.3} & 0.13^{+0.09}_{-0.03} & -1.45 & 0.09 & 1.53 \\
		2 & 17 & 1 & 124.2 \pm 2.4 & 599.4 \pm 275.4 & 4.8 \pm 2.2 & 0.7^{+0.8}_{-0.4} & 5.0^{+3.4}_{-2.9}  & 12.2^{+5.2}_{-5.9} & 0.13^{+0.09}_{-0.04} & -1.78 & 0.10 & 1.88 \\
		1 & 22 & \nodata & 436 \pm 4.5 & 2129 \pm 280 & 4.9 \pm 0.6 & 0.2^{+0.5}_{-0.1} & 3.9^{+4.0}_{-2.3}  & 18.4^{+1.2}_{-2.3} & 0.09^{+0.01}_{-0.00} & -0.88 & -0.37 & 0.51 \\
		1 & 23 & \nodata & 243 \pm 6.4 & 1200 & 4.9 \pm 0.1 & 0.2^{+0.5}_{-0.0} & 3.8^{+4.1}_{-2.3}  & 19.4^{+0.5}_{-0.6} & 0.09^{+0.00}_{-0.00} & -4.20 ^* & -0.38 & 3.82 ^* \\
		2 & 32 & 2 & 202.8 \pm 1.2 & 1146.6 \pm 291.0 & 5.7 \pm 1.4 & 0.3^{+0.5}_{-0.1} & 3.9^{+4.0}_{-2.4}  & 15.9^{+2.9}_{-4.5} & 0.09^{+0.02}_{-0.01} & -2.89 ^* & -0.35 & 2.53 ^* \\
		\enddata
		
		\tablecomments{Columns present the reference (Ref), the event ID (Ev.ID) and loop ID (Lp.ID) from references, the observed period (P), the damping time ($\tau_{\rm d}$), damping rate (r), the parameter inferred median values of the three selected theoretical models ($\zeta$, l/R, w, R/L) with their uncertainties, and Bayes factor. The symbol ($^*$) indicate Positive Evidence according to the corresponding Bayes factor.}
		\tablereferences{(1) \citet{Verwichte2013}; (2) \citet{Goddard2016} }
	\end{deluxetable}
\end{longrotatetable}

\end{document}